\documentclass[12pt,draftclsnofoot,onecolumn,peerreview]{IEEEtran}
\usepackage{graphicx}
\usepackage[cmex10]{amsmath}\interdisplaylinepenalty=2500
\usepackage{multirow}
\usepackage{url}
\usepackage{hyperref}
\usepackage[printonlyused,withpage]{acronym}
\usepackage{amsthm}
\usepackage{amssymb}
\usepackage{mathtools, cuted}\DeclarePairedDelimiter{\ceil}{\lceil}{\rceil}
\usepackage[monochrome]{color}
\usepackage{booktabs}
\usepackage{array}
\usepackage[usenames,dvipsnames]{xcolor}
\usepackage{setspace}

\newcommand{\lT}{\ensuremath{\lambda_\text{T}}}
\newcommand{\lI}{\ensuremath{\lambda_\text{I}}}
\newcommand{\lA}{\ensuremath{\lambda_\text{A}}}
\newcommand{\lR}{\ensuremath{\lambda_\text{R}}}
\newcommand{\lS}{\ensuremath{\lambda_\text{S}}}

\newlength{\figwidth}
\ifCLASSOPTIONtwocolumn
\fi
\ifCLASSOPTIONonecolumn
\fi

\title{Assessment of LTE Wireless Access for \\Monitoring of Energy Distribution \\ in the Smart Grid}
\author{\IEEEauthorblockN{Germ\'an~C.~Madue\~{n}o, Jimmy~J.~Nielsen, Dong~Min~Kim,  Nuno~K.~Pratas, \v Cedomir Stefanovi\'c, Petar~Popovski} \\
\IEEEauthorblockA{Department of Electronic Systems, Aalborg University, Denmark\\
Email: \{gco,jjn,dmk,nup,cs,petarp\}@es.aau.dk}%
}

\begin{document}
\maketitle

\begin{abstract}
While LTE is becoming widely rolled out for human-type services, it is also a promising solution for cost-efficient connectivity of the smart grid monitoring equipment.
This is a type of machine-to-machine (M2M) traffic that consists mainly of sporadic uplink transmissions. 
In such a setting, the amount of traffic that can be served in a cell is not constrained by the data capacity, but rather by the signaling constraints in the random access channel and control channel.
In this paper we explore these limitations using a detailed simulation of the LTE access reservation protocol (ARP).
We find that 1) assigning more random access opportunities may actually worsen performance; and 2) the additional signaling that follows the ARP has very large impact on the capacity in terms of the number of supported devices; we observed a reduction in the capacity by almost a factor of 3.
This suggests that a lightweight access method, with a reduced number of signaling messages, needs to be considered in standardization for M2M applications.
Additionally we propose a tractable analytical model to calculate the outage that can be rapidly implemented and evaluated.
The model accounts for the features of the random access, control channel and uplink and downlink data channels, as well as retransmissions.
\end{abstract}


\IEEEpeerreviewmaketitle

\section{Introduction} 
\label{sec:introduction}

\begin{figure}[tb]
	\centering
	\includegraphics[width=\figwidth]{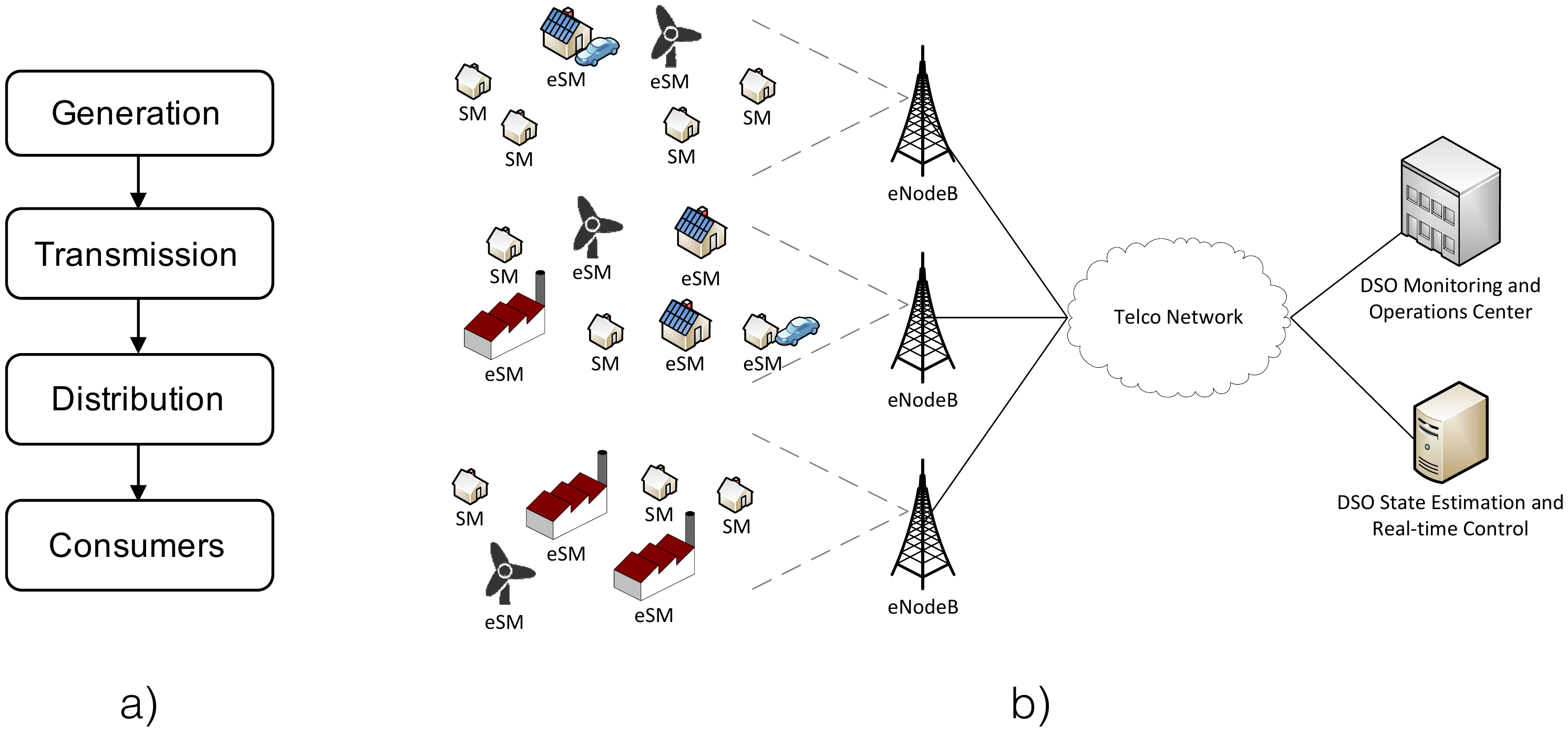}
	\caption{a) High level architecture of power grid. b) Cellular smart grid with smart meters (SM) and enhanced smart meters (eSM).}
	\label{fig:CellularSmartGrid}
\end{figure}

\textcolor{blue}{The defining feature of the evolution of traditional power grid toward smart grid is the inclusion of information and communication technologies in all segments of the power grid.
Fig.~\ref{fig:CellularSmartGrid} depicts a high-level diagram of power grid architecture; currently the communications for monitoring and control are widely used in generation and transmission domain, in the form of the wide area measurement systems (WAMS).
In addition, we are currently witnessing extensive deployments of the smart meters (SMs), i.e., network-connected electricity meters in the consumers domain, 
primarily used by electricity providers for availability monitoring and billing.}

\textcolor{blue}{On the other hand, in the distribution domain, the distribution system operators (DSOs) rely mainly on open loop control methods, i.e., there is no real-time monitoring and control in place, and the distribution grid is yet to be integrated in the smart grid monitoring and control framework.
One of the main drivers for the advanced monitoring and control of the distribution grid is the increasing penetration of distributed energy sources (DERs), and the roll-out of charging stations for electric vehicles.
Specifically, the integration of these novel power-grid elements into the distribution grid introduces highly variable and unpredictable variations in the power quality, requiring tighter monitoring and control.
To achieve this, DSOs will have to retrieve frequently updated measurements/samples at key points in the distribution grid.
This type of augmented observability of the distribution grid will be enabled by an advanced monitoring node, 
denoted in further text as an \emph{enhanced smart meter (eSM)}.
The eSM capabilities are expected to be similar to the ones currently available on a WAMS node, i.e., it should have Phasor Measurement Unit (PMU)-like capabilities.
This will allow eSMs to measure power quality parameters (such as power phasors) more frequently and in more detail compared to SMs~\cite{ETSI+TR102.935.2012}.
The fraction of eSMs needed in the distribution grid to achieve satisfactory state estimation is still an open research question \cite{huang2012state} and will have a definite impact on the requirements of the communication network that will provide connectivity.}

%

\textcolor{blue}{The above described monitoring and control of the distribution grid is an example of  Machine-to-Machine (M2M) communication, and, in broader context, the Internet of Things (IoT). 
\textcolor{blue}{Currently, there are several competing approaches that deal with the efficient provision of network access for M2M applications, relying on proprietary and open-standard technologies, e.g., LoRa \cite{lora}, SigFox \cite{sigfox}, IEEE 802.11ah \cite{qualcommProjects}, or cellular networks \cite{3gpp}.}
In this paper we focus on the latter, i.e., on the use of cellular technologies and investigate the usability of an LTE access network to support monitoring applications in the smart distribution grid.
The motivation arises naturally from: (a) the expected ubiquitous presence and advanced capabilities of LTE, (b) the savings of the capital and operational expenses that DSOs may expect when using the existing telco infrastructure, and (c) the incentive for the telcos to support smart grid applications, which are seen as new sources of revenue.
The presented study focuses on the operation of the LTE access protocol and, in contrast to the existing works, takes into account all the aspects that influence its operation when supporting the potentially large number of eSMs within an LTE cell.
Specifically, we present a thorough analysis of the LTE access protocol that includes all signaling overheads, investigate its performance and limitations under distribution grid monitoring scenarios, and draw important conclusions with respect to the dimensioning and resource allocation of the access mechanism.
To the best of our knowledge, there is no such study in the previous literature. 
Our ultimate goal is to provide the standardization bodies and mobile operators with insights that can influence the relevant M2M standardization activities and the M2M-oriented evolution of the cellular networks.}

\textcolor{blue}{
The rest of this paper is organized as follows.
We begin with a detailed description of LTE access reservation procedure in Section \ref{sec:lte_overview}.
In Section~\ref{sec:bg_rw_contrib} we provide the motivation of this work, review the 
relevant previous works, and outline the contributions of the paper.
The analytical model of LTE access procedure, which is the pivotal part of the paper, is provided in Section~\ref{sec:arp_model}.
 In Section \ref{sec:results} we present numerical
results, where the performance figures obtained with the proposed analytical model are compared to the ones obtained by simulation.
The conclusions are given in Section \ref{sec:conclusions}.}

\textcolor{blue}{We conclude this section by listing in Table~\ref{tab:acronyms} the acronyms that are used throughout the paper.}
	\begin{table}[tb]
	\centering\footnotesize
	\caption{Acronyms List}
	\label{tab:acronyms}
	\begin{tabular}{|c|l|}
	\hline
	\textbf{Acronym} & \textbf{Description} \\ \hline
	ARP & Access Reservation Procedure\\
	CCE & Control Channel Element\\
	CFI & Control Format Indicator\\
	DER  & Distributed Energy source\\
	DSO  &  Distribution System Operators\\ 
	eSM  &  Enhanced Smart Meter\\
	NAS & Network Access Stratum\\
	PBCH &  Physical Broadcast Channel\\
	PCFICH & Physical Control Format Indicator Channel\\
	PDCCH &  Physical Downlink Control Channel\\
	PDSCH &  Physical Downlink Shared Channel\\
	PHICH &  Physical Hybrid ARQ Channel\\
	PMU  &  Phasor Measurement Unit\\
	PSS  & Primary Synchronization Signal\\
	PUCCH  &  Physical Uplink Control Channel\\
	PUSCH  &  Physical Uplink Shared Channel\\
	RAR  &  Random Access Response (MSG~2)\\
	RB  &  Resource Block\\
	RE & Resource Element \\
	RRC  & Radio Resource Control\\
	SM  &  Smart Meter\\
	SSS  & Secondary Synchronization Signal\\
	WAMS  & Wide Area Measurement Systems\\
	\hline
	\end{tabular}
\end{table}


\section{Detailed Description of LTE Access}
\label{sec:lte_overview}


	In this section, we first describe the organization of the LTE access resources and channel in the downlink and uplink.
	We then turn to the description of the connection establishment.
	

	\subsection{Downlink} 
	\label{sub:downlink}
	
\begin{figure}[tb]
	    \centering
	    \includegraphics[width=\figwidth]{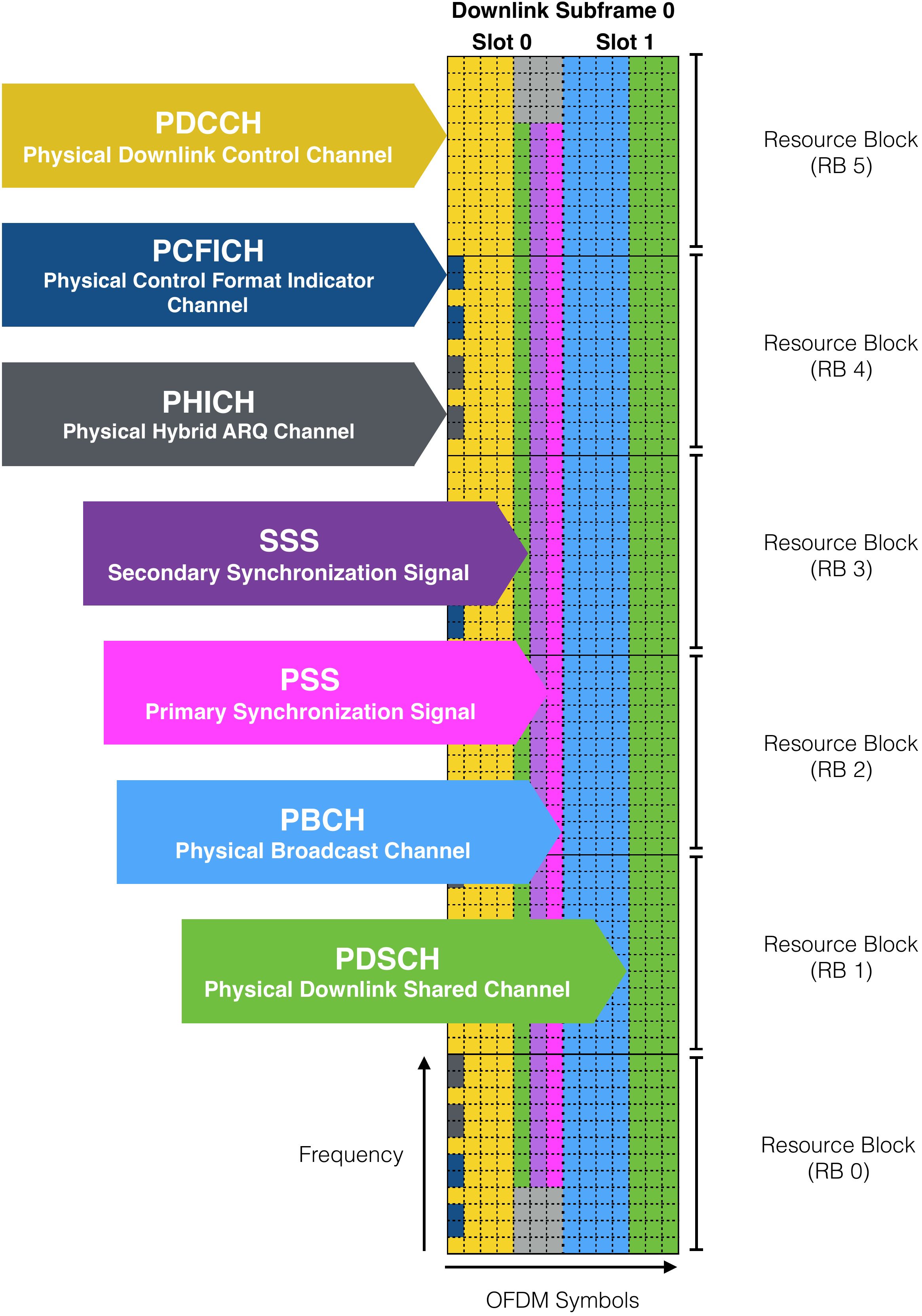}
	    \caption{Simplified illustration of downlink subframe 0 organization in a 1.4~MHz system with $N_\text{CFI}=3$.}
	    \label{fig:downlinkGrid}
	\end{figure}

	The downlink resources in LTE in the case of frequency division duplexing (FDD) are divided into time-frequency units, where the smallest unit is denoted as a \emph{resource element (RE)}.
	 Specifically, the time is divided in frames, where every frame has ten subframes, and each subframe is of duration $t_s = 1 \, \text{ms}$.
	An illustration of a subframe is presented in Fig.~\ref{fig:downlinkGrid}.
	Each subframe is composed in time by 14 OFDM modulated symbols, where the amount of bits of each symbol depends on the modulation used, which could be QPSK, 16QAM or 64QAM.
	The system bandwidth determines the number of frequency units available in each subframe, which is typically measured in resource blocks (RBs), where a RB is composed by 12 frequency units and 14 symbols, i.e., a total of 168 REs.
	The amount of RBs in the system varies from 6~RBs in 1.4~MHz system to 100~RBs in 20~MHz system.
	
	\begin{table}[tb]
		\centering\footnotesize
		\caption{PDCCH Formats in LTE}
		\label{table:pddchFormats}
		\begin{tabular}{|c|l|c|}
		\hline
		{\bf Format} & {\bf Purpose}                             & {\bf No. of CCEs} \\ \hline 
		0            & Transmission of resource grants for PUSCH & 1                    \\ \hline
		1            & Scheduling PDSCH                          & 2                    \\ \hline
		2            & Same as 1 but with MIMO                   & 4                    \\ \hline
		3            & Transmission of power control commands    & 8                    \\ \hline
		\end{tabular}
	\end{table}

	\begin{table}[tb]
		\centering\footnotesize
		\caption{Number of CCEs per subframe}
		\label{table:numPDCCHs}
		\begin{tabular}{|c|c|c|c|}
		\hline
		\multirow{2}{*}{\bf System Bandwidth} &
			\multicolumn{3}{c|}{\bf Number of CCEs} \\ \cline{2-4} 
			& $CFI=1$ & $CFI=2$ & $CFI=3$ \\ \hline
		1.4~MHz 	& 2 	& 4 	& 6 \\ \hline
		5~MHz 		& 4 	& 13 	& 21 \\ \hline
		10~MHz 		& 10 	& 26 	& 43 \\ \hline
		20~MHz 		& 20 	& 54 	& 87 \\ \hline
		\end{tabular}
	\end{table}

	\textcolor{blue}{In the downlink, there are two main channels; these are the physical downlink control channel (PDCCH) and the physical downlink shared channel (PDSCH).
	The PDCCH carries the information about the signaling/data being transmitted on the current PDSCH and the information about the resources which the devices need to use for the physical uplink shared channel (PUSCH), as illustrated in Fig.~\ref{fig:downlinkGrid}.
	Therefore, signaling and data messages consume resources both in the control and shared data channels.}
	The PDCCH is composed by the first $N_\text{CFI}$ symbols in each subframe. This value is controlled by the CFI parameter indicated in the physical control format indicator channel (PCFICH)~\cite{3GPPTS36.212}, see Fig.~\ref{fig:downlinkGrid}.\footnote{Note that not all REs are used for PDCCH, some of them are reserved for other channels such as the PCFICH and the physical hybrid indicator channel (PHICH).}
	The CFI takes values $N_\text{CFI}=\{1, 2$,~or~$3\}$, where it is recommended to use $N_\text{CFI}=3$ for a system bandwidth of 1.4~MHz and 5~MHz and $N_\text{CFI}=2$ for a system bandwidth of 10~MHz to 20~MHz \cite{3GPPTS36.508}.
	It should be noted that 1.4~MHz is a special case, where $N_\text{CFI}=1$ dedicates the first two symbols for PDCCH and $N_\text{CFI}=3$ the first four symbols.
	The amount of PDCCH resources taken for every message, which is measured in control channel elements (CCEs), depends on the PDDCH format required for the type of MAC message the eNodeB wishes to transmit.
	A CCE is composed by 36 REs, and there are four formats of PDCCS available in LTE-A, listed in Table~\ref{table:pddchFormats} together with the amount of CCE required.
	For the sake of simplicity, we focus on PDCCH format $1$, which is the one used for the described messages, especially in the case of M2M with no MIMO capabilities \cite{3GPPTS36.523-3}. When format 1 with 2 CCEs is used, the maximum number of PDCCH messages per subframe in a 1.4~MHz system system is three \cite{3GPPTS36.523-3}. This emphasizes the importance of modeling the limitations imposed by the PDCCH. 

	The remaining resources are used for the physical broadcast channel (PBCH), primary and secondary synchronization signals (PSS and SSS respectively), and PDSCH, as shown in Fig.~\ref{fig:downlinkGrid}.\footnote{We note that PSS and SSS only take place every 5 subframes.}
	Obviously, there is a scarcity of resources for MAC messages in the PDSCH.	


	\subsection{Uplink} 
	\label{sub:uplink}
	
\begin{figure}[tb]
	    \centering
	    \includegraphics[width=\figwidth]{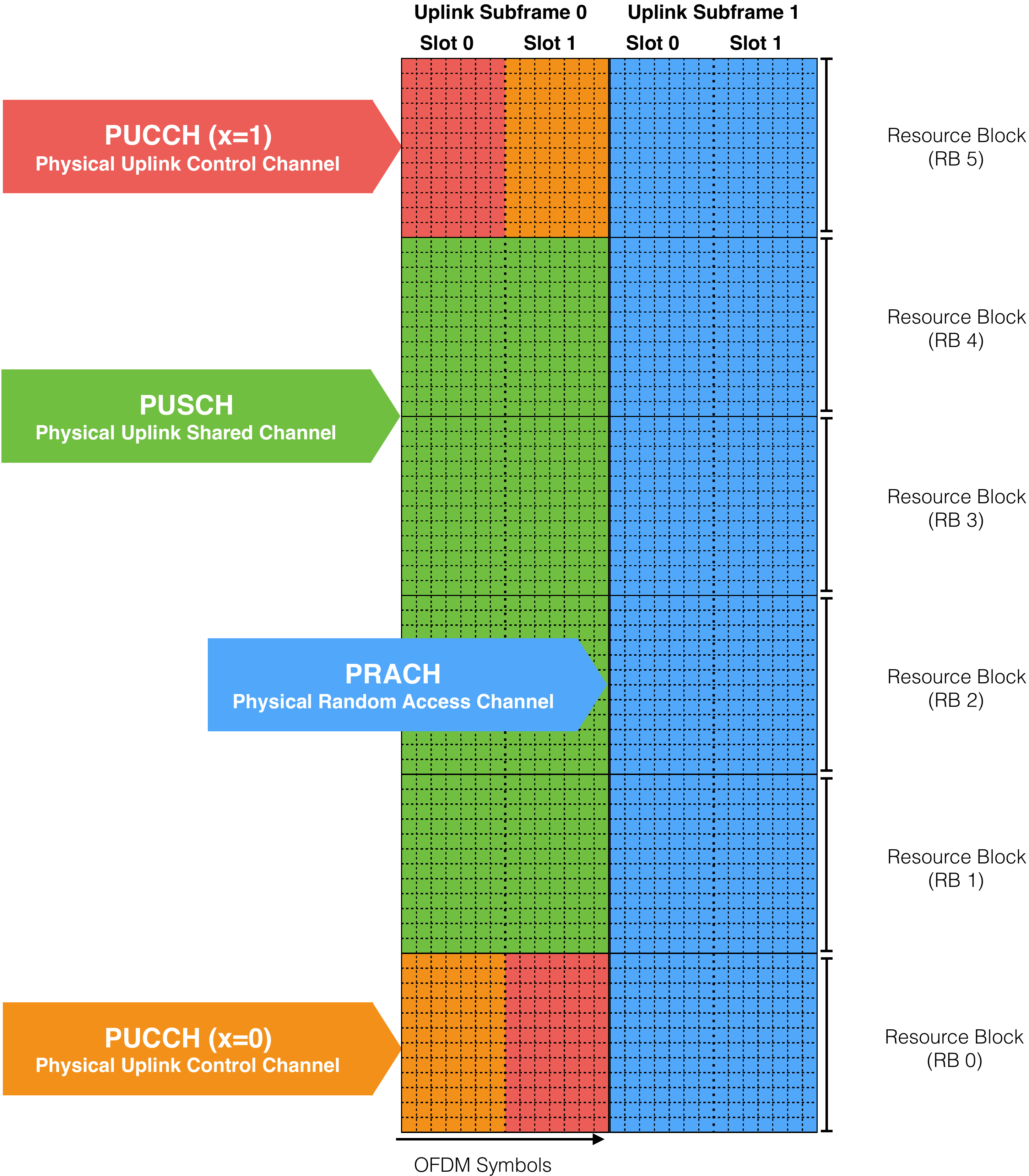}
	    \caption{Simplified illustration of uplink subframe 0 and subframe 1 organization in a 1.4~MHz system with $N_\text{CFI}=3$.}
	    \label{fig:uplinkGrid}
	\end{figure}

	The uplink resources are organized similarly as in the downlink, with the main difference that the smallest resource that can be addressed is a RB.
	\textcolor{blue}{The physical uplink shared channel (PUSCH) is used by devices for signaling and data messages, where it should be noted that several devices can be multiplexed on the same subframe.}
	As shown in Fig.~\ref{fig:uplinkGrid}, the physical uplink control channel (PUCCH) takes place in RB~0 in slot 0 and then in RB~5 in slot 1 (x=0), where $x$ denotes if the PUCCH index.\footnote{PUCCH Index is used to indicate to user which PUCCH resources shall be used.}
	In order words, to enable frequency diversity the PUCCH transmission takes place in the lowest and highest part of the frequency grid.

	When present the PRACH occupies 6 RBs and occurs periodically, from once in every two frames (20 sub-frames) to once in every sub-frame.
	A typical PRACH periodicity value is once every 5 sub-frames~\cite{typical}.
	%


	\subsection{LTE Connection Establishment} 
	\label{sub:lte_access_reservation_protocol}

\begin{figure}[h]
    \centering
    \includegraphics[width=\figwidth]{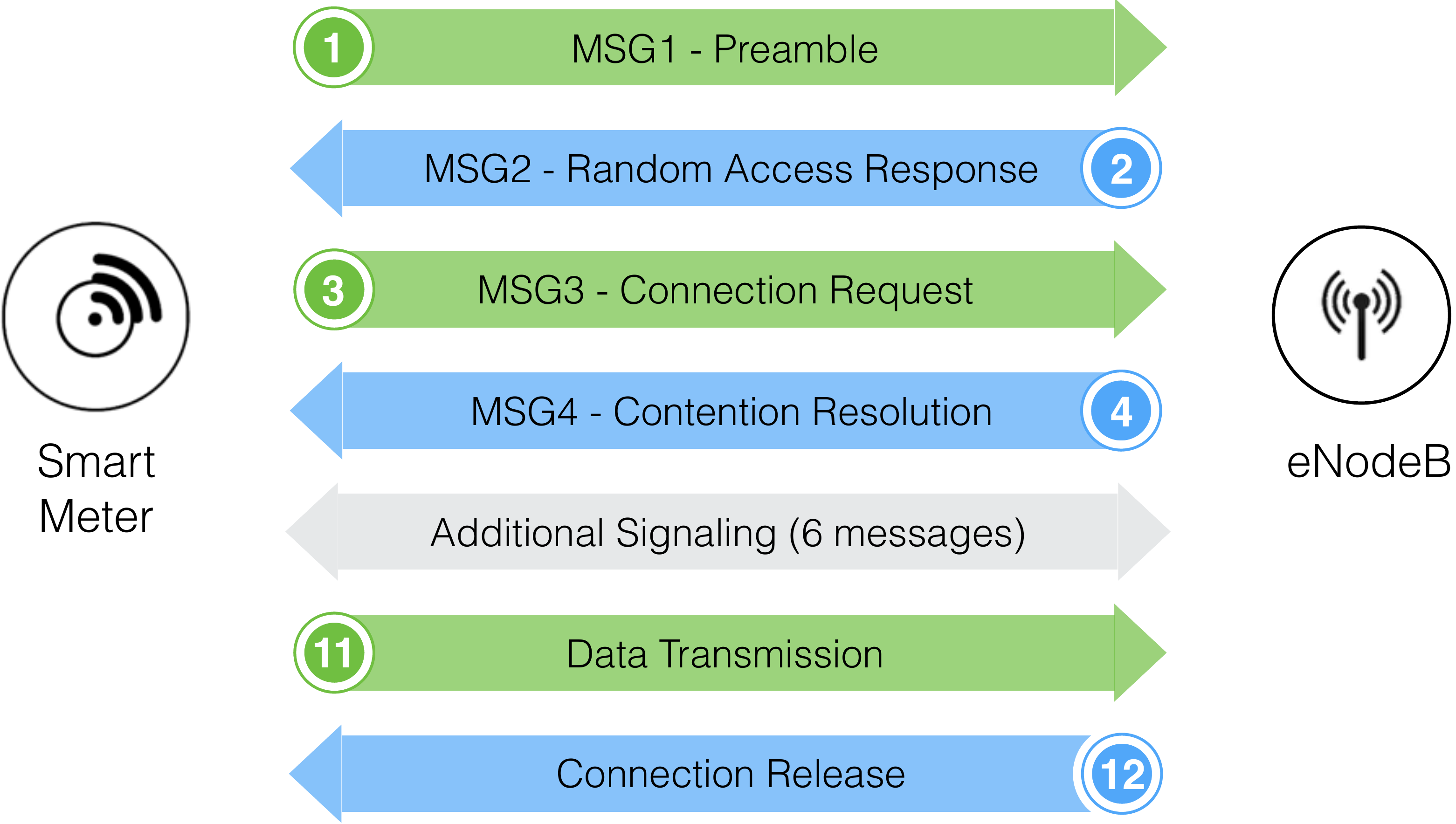}
    \caption{Message exchange between a smart meter and the eNodeB.}
    \label{fig:LTE_msg1-msg4}
\end{figure}

		The connection establishment in LTE starts with the access reservation procedure.
		The ARP in LTE consists of the exchange of four MAC messages between the accessing device, in further text denoted as user equipment (UE), and the eNodeB, as shown in Fig.~\ref{fig:LTE_msg1-msg4}.
		The first message (MSG~1) is a random access preamble sent in the first random access opportunity (RAO) that is available,
		where RAO is a PRACH subframe.
		The number of subframes between two RAOs varies between 1 and 20, and it is denoted as $\delta_\text{RAO}$.
		In other words, $\delta_\text{RAO}$ indicates the number of subframes between PRACH occurrences.
		The preambles that UEs contend with are randomly chosen from the set of 64 orthogonal preambles, where only $d=54$ are typically available for contention purposes and the rest are reserved for timing alignment.
		The contention is slotted ALOHA based~\cite{3GPPTS36.321,3GPPTS36.213}, but unlike in typical ALOHA scenarios, the eNodeB can only detect which preambles have been activated but not if multiple activations (collisions) have occurred.
		In particular, this assumption holds in small/urban cells~\cite[Sec. 17.5.2.3]{sesia2011lte}.\footnote{If the cell size is more than twice the distance corresponding to the maximum delay spread, the eNodeB may be able to differentiate the case that preamble has been activated by two or more users, but \emph{only} if the users are separable in terms of the Power Delay Profile \cite{sesia2011lte,6530825}.}
		
		Via MSG~2, the eNodeB returns a random access response (RAR) to all detected preambles.
		The contending devices listen to the downlink channel, expecting MSG~2 within time period $t_{\mathrm{RAR}}$.
		If no MSG~2 is received and the maximum of $T$ MSG~1 transmissions has not been reached, the device backs off and restarts the random access procedure after a randomly selected backoff interval $t_r\in[0,W_\text{c}-1]$. 
		If received, MSG~2 includes uplink grant information that indicates the RB in which the connection request (MSG~3) should be sent.
		The connection request specifies the requested service type, e.g., voice call, data transmission, measurement report, etc.
		When two devices select the same preamble (MSG~1), they receive the same MSG~2 and experience collision when they send their MSG~3s in the same RB.

		In contrast to the collisions for MSG~1, the eNodeB is able to detect collisions for MSG~3. The eNodeB only replies to the MSG~3s that did not experience collision, by sending message MSG~4 (i.e., RRC Connection Setup). The message MSG~4 may carry two different outcomes: either the required RBs are allocated or the request is denied in case of insufficient network resources. The latter is however unlikely in the case of M2M communications, due to the small payloads. If the MSG~4 is not received within time period $t_{\mathrm{CRT}}$ since MSG~1 was sent, the random access procedure is restarted. Finally, if a device does not successfully finish all the steps of the random access procedure within $m\!+\!1$ MSG~1 transmissions, an outage is declared.
		


	 \setlength\tabcolsep{3 pt}
 {\renewcommand{\arraystretch}{1.3}%
\begin{table*}[t]
  \centering\footnotesize
  \caption{List of messages exchanged between the smart meter and the eNodeB.}
    \begin{tabular}{|c|c|c|l|c|c|}

    \hline
    & \multirow{2}{*}{\bf Step} & \multirow{2}{*}{\bf Channel}  & \multirow{2}{*}{\bf Message}  & \multicolumn{2}{c|}{\bf MAC Size (Bytes)} \\ \cline{5-6}
		& & & 	& Uplink & Downlink \\ \hline

    \parbox[t]{2mm}{\multirow{8}{*}{\rotatebox[origin=c]{90}{ARP}}}
    &\textcolor{black}{ 1     }&\textcolor{black}{ $\uparrow$    PRACH }&\textcolor{black}{ MSG 1: Preamble }&\textcolor{black}{ --    }&\textcolor{black}{ -- }\\
    &\textcolor{gray}{ 2     }&\textcolor{gray}{ $\downarrow$  PDCCH }&\textcolor{gray}{ \textcolor{gray}{Downlink Grant} }&\textcolor{gray}{ --    }&\textcolor{gray}{ -- }\\
    &\textcolor{black}{ 3     }&\textcolor{black}{ $\downarrow$  PDSCH }&\textcolor{black}{ MSG 2: Random Access Response }&\textcolor{black}{ --    }&\textcolor{black}{ 8 }\\
    &\textcolor{black}{ 4     }&\textcolor{black}{ $\uparrow$    PUSCH }&\textcolor{black}{ MSG 3: RRC Connection Request }&\textcolor{black}{ \textbf{7} }&\textcolor{black}{ \textbf{--} }\\
    &\textcolor{gray}{ 5     }&\textcolor{gray}{ $\downarrow$  PHICH }&\textcolor{gray}{ ACK   }&\textcolor{gray}{ --    }&\textcolor{gray}{ -- }\\
    &\textcolor{gray}{ 6     }&\textcolor{gray}{ $\downarrow$  PDCCH }&\textcolor{gray}{ Downlink Grant }&\textcolor{gray}{ --    }&\textcolor{gray}{ -- }\\
    &\textcolor{black}{ 7     }&\textcolor{black}{ $\downarrow$  PDSCH }&\textcolor{black}{ MSG 4: RRC Connection Setup }&\textcolor{black}{ --    }&\textcolor{black}{ 38 }\\
    &\textcolor{gray}{ 8     }&\textcolor{gray}{ $\downarrow$  PUCCH }&\textcolor{gray}{ ACK   }&\textcolor{gray}{ --    }&\textcolor{gray}{ -- }\\

    \hline

    \parbox[t]{2mm}{\multirow{17}{*}{\rotatebox[origin=|c|]{90}{Additional Signaling}}}
    &\textcolor{gray}{ 9     }&\textcolor{gray}{ $\downarrow$  PUCCH }&\textcolor{gray}{ Scheduling Request }&\textcolor{gray}{ --    }&\textcolor{gray}{ -- }\\
    &\textcolor{gray}{ 10    }&\textcolor{gray}{ $\downarrow$  PDCCH }&\textcolor{gray}{ UL Grant }&\textcolor{gray}{ --    }&\textcolor{gray}{ -- }\\
    &\textcolor{black}{ 11    }&\textcolor{black}{ $\downarrow$  PUSCH }&\textcolor{black}{ RRC Connection Setup Complete (+NAS: Service Req. and Buffer Status) }&\textcolor{black}{ 20    }&\textcolor{black}{ -- }\\
    &\textcolor{gray}{ 12    }&\textcolor{gray}{ $\downarrow$  PHICH }&\textcolor{gray}{ ACK   }&\textcolor{gray}{ --    }&\textcolor{gray}{ -- }\\
    &\textcolor{gray}{ 13    }&\textcolor{gray}{ $\downarrow$  PDCCH }&\textcolor{gray}{ Downlink Grant }&\textcolor{gray}{ --    }&\textcolor{gray}{ -- }\\
    &\textcolor{black}{ 14    }&\textcolor{black}{ $\downarrow$  PDSCH }&\textcolor{black}{ Security Mode Command }&\textcolor{black}{ --    }&\textcolor{black}{ 11 }\\
    &\textcolor{gray}{ 15    }&\textcolor{gray}{ $\uparrow$    PUCCH }&\textcolor{gray}{ ACK   }&\textcolor{gray}{ --    }&\textcolor{gray}{ -- }\\
    &\textcolor{gray}{ 19    }&\textcolor{gray}{ $\uparrow$    PUCCH }&\textcolor{gray}{ Scheduling Request }&\textcolor{gray}{ --    }&\textcolor{gray}{ -- }\\
    &\textcolor{gray}{ 20    }&\textcolor{gray}{ $\downarrow$  PDCCH }&\textcolor{gray}{ UL Grant }&\textcolor{gray}{ --    }&\textcolor{gray}{ -- }\\
    &\textcolor{black}{ 21    }&\textcolor{black}{ $\uparrow$    PUSCH }&\textcolor{black}{ Security Mode Complete }&\textcolor{black}{ 13    }&\textcolor{black}{ -- }\\
    &\textcolor{gray}{ 22    }&\textcolor{gray}{ $\downarrow$  PHICH }&\textcolor{gray}{ ACK   }&\textcolor{gray}{ --    }&\textcolor{gray}{ -- }\\
    &\textcolor{gray}{ 16    }&\textcolor{gray}{ $\downarrow$  PDCCH }&\textcolor{gray}{ Downlink Grant }&\textcolor{gray}{ --    }&\textcolor{gray}{ -- }\\
    &\textcolor{black}{ 17    }&\textcolor{black}{ $\downarrow$  PDSCH }&\textcolor{black}{ RRC Connection Reconfiguration (+NAS: Activate EPS Bearer Context Req.) }&\textcolor{black}{ --    }&\textcolor{black}{ 118 }\\
    &\textcolor{gray}{ 18    }&\textcolor{gray}{ $\uparrow$    PUCCH }&\textcolor{gray}{ ACK   }&\textcolor{gray}{ --    }&\textcolor{gray}{ -- }\\
    &\textcolor{gray}{ 23    }&\textcolor{gray}{ $\uparrow$    PUCCH }&\textcolor{gray}{ Scheduling Request }&\textcolor{gray}{ --    }&\textcolor{gray}{ -- }\\
    &\textcolor{gray}{ 24    }&\textcolor{gray}{ $\downarrow$  PDCCH }&\textcolor{gray}{ UL Grant }&\textcolor{gray}{ --    }&\textcolor{gray}{ -- }\\
    &\textcolor{black}{ 25    }&\textcolor{black}{ $\uparrow$    PUSCH }&\textcolor{black}{ RRC Connection Reconfiguration Complete  }&\textcolor{black}{ 10    }&\textcolor{black}{ -- }\\
    &\textcolor{gray}{ 26    }&\textcolor{gray}{ $\downarrow$  PHICH }&\textcolor{gray}{ ACK   }&\textcolor{gray}{ --    }&\textcolor{gray}{ -- }\\

    \hline

    \parbox[t]{2mm}{\multirow{4}{*}{\rotatebox[origin=c]{90}{Data}}}
    &\textcolor{gray}{ 27    }&\textcolor{gray}{ $\uparrow$    PUCCH }&\textcolor{gray}{ Scheduling Request }&\textcolor{gray}{ --    }&\textcolor{gray}{ -- }\\
    &\textcolor{gray}{ 28    }&\textcolor{gray}{ $\downarrow$  PDCCH }&\textcolor{gray}{ UL Grant }&\textcolor{gray}{ --    }&\textcolor{gray}{ -- }\\
    &\textcolor{black}{ 29    }&\textcolor{black}{ $\uparrow$    PUSCH }&\textcolor{black}{ Report (Data) }&\textcolor{black}{ Variable }&\textcolor{black}{ \textbf{--} }\\
    &\textcolor{gray}{ 30    }&\textcolor{gray}{ $\downarrow$  PHICH }&\textcolor{gray}{ ACK   }&\textcolor{gray}{ --    }&\textcolor{gray}{ -- }\\

    \hline

    &\textcolor{gray}{ 31    }&\textcolor{gray}{ $\downarrow$  PDCCH }&\textcolor{gray}{ Downlink Grant }&\textcolor{gray}{ --    }&\textcolor{gray}{ -- }\\
    &\textcolor{black}{ 32    }&\textcolor{black}{ $\downarrow$  PDSCH }&\textcolor{black}{ RRC Connection Release }&\textcolor{black}{ --    }&\textcolor{black}{ 10 }\\
    \hline
    \end{tabular}%
  \label{tab:signalingMessages}%
\end{table*}%
 \setlength\tabcolsep{6 pt}

	After ARP exchange finishes, there is an additional exchange of MAC messages between the smart meter and the eNodeB, whose main purposes is to establish security and quality of service for the connection, as well as to indicate the status of the buffer at the device. These extra messages are detailed further in Table~\ref{tab:signalingMessages}. 
	
	Besides MAC messages, there are PHY messages included in the connection establishment \cite{signaling3GPP}.
	Table~\ref{tab:signalingMessages} presents a complete account of both PHY and MAC messages exchanged during connection establishment, data report transmission and connection termination (the PHY messages are indicated in gray).
	As it can be seen from the table, for every downlink message a downlink grant in the PDCCH is required.
	Similarly, every time a smart meter wishes to transmit in the uplink after the ARP, it first need to ask for the uplink resources by transmitting a scheduling request in the PUCCH.\footnote{We note that the amount of resources reserved for PUCCH is very small for scheduling periodicity above 40~ms \cite{signaling3GPP} and therefore will not be considered in the following text and analysis.}
	This is followed by provision of an uplink grant in the PDCCH by the eNodeB.


\section{Motivation, Related Work and Contributions}
\label{sec:bg_rw_contrib}

As already outlined, the traffic profile generated by smart-grid monitoring devices is
an example of Machine-to-Machine (M2M) traffic, characterized by a
sporadic transmissions of small amounts of data from a very large number
of terminals. This is in sharp contrast with the bursty and
high data-rate traffic patterns of the human-centered services.
\textcolor{blue}{
		Another important difference is that smart grid services typically require a higher degree of network reliability and availability than the human-centered services \cite{6577600}. So far, cellular access has been optimized to human-centered traffic and M2M related standardization efforts came into focus only recently~\cite{3GPPR11}.
	}

Due to the sporadic, i.e., intermittent nature of M2M communications, it is typically assumed that the M2M devices will have to establish the connection to the cellular access network every time they perform reporting.
From Section~\ref{sec:lte_overview} it becomes apparent that connection establishment requires extensive signaling, both in the uplink and the downlink, and the total amount of the  signaling information that is exchanged may well over outweigh the information contained in the data report.
Moreover, the total number of resources available in the uplink and downlink is limited, and in the case of a massive number of M2M devices, the signaling traffic related to the establishment of many connections may pose a significant burden to the operation of the access protocol.
Thus, it is of paramount importance to consider the whole procedure associated with the transmission of a data (report) in order to properly estimate the number of M2M devices that can be supported in the LTE access network.



	\subsection{Related Work} 
	\label{sub:related_work}
	
	Simple models to determine the probability of  preamble collision (MSG~1) in the PRACH channel are presented in 3GPP standard documents \cite{3gpp2006R1-061369,3gpp2011R2-112198,3gpp2011tr37868} and in the scientific literature \cite{cheng2012rach,ubeda2012lte,karupongsiri2014random}.
	\textcolor{blue}{Reception of a preamble is based on energy detection~\cite{6824261} and a detected preamble indicates that there is at least one active user that sends that preamble. The drawback is the inability of the receiver to discern if a preamble has been selected just by a single device or by multiple devices~\cite{6530825}.}
	More specifically, the eNodeB can only infer whether the preamble is activated, but not how many devices have simultaneously activated it.

	\textcolor{blue}{To alleviate the PRACH overload, a group paging is proposed \cite{wei2013performance}, where the base station adjusts the group size to prevent preamble collisions and PDCCH limitation.
	A related analytical model to represent the number of contending, failed, and success uplink attempts was developed, however, the effect of PDCCH resource limitation has not been taken into account.}
	\textcolor{blue}{An investigation of the ARP performance considering the effect of the limitation of PDCCH resources, by modeling the sharing of the PDCCH between MSG~2 and MSG~4 with priority placed on MSG~2 \cite{3gpp2011tr37868}, shows that the ARP performance is severely degraded when the LTE system accepts a large number of uplink devices during the second step of the PRACH procedure \cite{osti2014analysis}, which is due to the lazy handling of MSG~4. It was assumed that all uplink requests, including retransmissions, constitute a Poisson process, and evidence for this is provided via simulations.}
	\textcolor{blue}{The PDCCH sharing problem is raised in \cite{lin2015estimation}, and the PDCCH resource scheduling policy based on the solution of ARP throughput maximization problem is proposed.
	The authors also propose a dynamic backoff scheme as a remedy for the PRACH overload.}	
    In this paper, we present a more accurate analytical model compared to \cite{wei2013performance}-\cite{lin2015estimation}, as we are considering the effect of PDSCH and PUSCH limitations as well as the effect of PDCCH limitations.
    We also present a tractable model of the retransmission behavior of the uplink devices during the whole ARP.

	In the context of smart grid monitoring applications, a simplified evaluation of the cellular access performance, which neglects the impact of ARP, is performed in \cite{hagerling2014coverage,nist2011pap2}.
	However, it was shown that large differences in the performance of the network can be observed if the ARP is not considered \cite{nielsen2015magazine} , motivating the detailed study presented in this paper.
\textcolor{blue}{Specifically, in \cite{nielsen2015magazine} we investigate and specify smart meter traffic models and present a simulation-based study of the ARP limitations; however the model extents only up to MSG~4.
We also note that the analytical model and simulation framework used in this paper are more detailed versions of the preliminary material presented in \cite{nielsen2015tractable}.
The main difference is that the analysis in \cite{nielsen2015tractable} does not consider a detailed modeling of the  PDCCH, PDSCH and PUSCH, but uses only a simple limitation on the number of uplink grants allowed per random access response (RAR) message. These simplifications are removed in the analysis performed in the present paper.} 
	

	\subsection{Our Contributions} 
	\label{sub:our_contributions}

	The contributions of the work presented here are:
	\begin{itemize}
	 \item Comprehensive study of the connection establishment between a device and eNodeB in the LTE context, which considers (i) both the uplink and downlink exchanges, and (ii) both PHY- and MAC-layer aspects.
	 \item Identification and modeling of the limitations of connection establishment. Specifically, we develop an analytical model that describes PRACH, PDCCH, PDSCH and PUSCH limitations. We show in the paper that the capacity of the access is decreased by a factor of almost three when these limitations are taken into account, in comparison to the studies that neglect them.
	 \item Development of a tractable model that describes the operation of the devices during the ARP. In order to fully characterize the ARP performance, we take into account a a retransmission strategy for the devices that do not successfully finish the ARP.
	 \item Based on the performed evaluation, we provide guidelines to future development of LTE in order to efficiently embrace traffic from the smart grid or similar M2M applications.
	\end{itemize}


	\begin{table}[h]
\centering\footnotesize
\label{tab:model_parameters}
\caption{Parameters used for the one-shot and $m$-retransmissions analytic models}
\begin{tabular}{|c|l|}
\hline
\textbf{Parameter} & \textbf{Description} \\ \hline
$N$ & Number of UEs in cell \\
$\lambda_\text{app}$ & Message generation rate per UE [msg/subframe]\\
$\lI$ & Message generation rate of all UEs in cell [msg/subframe]\\
$\lT$ & Access attempt rate of all UEs in cell including retransmission attempts [attempts/subframe]\\
$\lA$ & Mean number of activated preambles in cell [activations/subframe]\\
$\lR$ & Mean number of failed transmit attempts that lead to retransmissions in cell [failures/subframe]\\
$\lS$ & Mean number of singleton (non-collided) preambles in cell [singletons/subframe]\\
$m$ & Number of allowed retransmission attempts per message\\
$p_\text{f}$ & Probability of transmission attempt failing\\
$p_\text{c}$ & Probability of an activated preamble being involved in a collision\\
$p_\text{e}$ & Probability of failed connect request due to insufficient resources in PDCCH, PDSCH, or PUSCH\\
$d$ & Number of available preambles \\
$\delta_\text{RAO}$ & Interval between RAOs [subframes] \\
$\lambda$ & Arrival rate of requests to PDCCH, PDSCH, or PUSCH [requests/subframe]\\
$\mu$ & Service rate of PDCCH, PDSCH, or PUSCH [requests/subframe]\\
$\rho$ & Queue utilization factor\\
$p_\text{q}$ & Probability of M/M/1 queue not serving a message within deadline $T_\text{d}$ given $\lambda$ and $\mu$ parameters\\
$T_\text{d}$ & Deadline for serving an allocation request in PDCCH, PDSCH, or PUSCH\\
$p_\text{on}$ & Traffic generation probability\\
$W_\text{c}$ & Maximum backoff window size\\
$CR(i)$ & connection request state $i$\\
$P_\text{outage}$ & Probability of failing to deliver a message after up to $m$ retransmissions\\
$b_*$ & Steady-state probability of a given state *\\
$N_\text{TX}$ & Estimated average number of needed retransmissions\\
\hline
\end{tabular}
\end{table}

\section{Analysis}
\label{sec:arp_model} 

\textcolor{blue}{
		For simplicity we assume a single LTE cell with $N$ UEs. However, it should be noted the proposed model could be easily adapted to a more realistic scenario with inter-cell interference as the main difference would be a decreased packet transmission success probability, mainly due to a lower SNIR.
		} Further, we assume that the smart grid application, associated with UEs, generates new uplink transmissions with an
aggregate rate that is Poisson distributed with parameter $\lI$, as depicted in Fig.~\ref{fig:LTE_diagram}; 
\textcolor{blue}{note that the unit of $\lambda_\text{I}$ is the number of transmission attempts per second.}
\begin{figure*}[bt]
    \centering
    \includegraphics[width=0.7\linewidth]{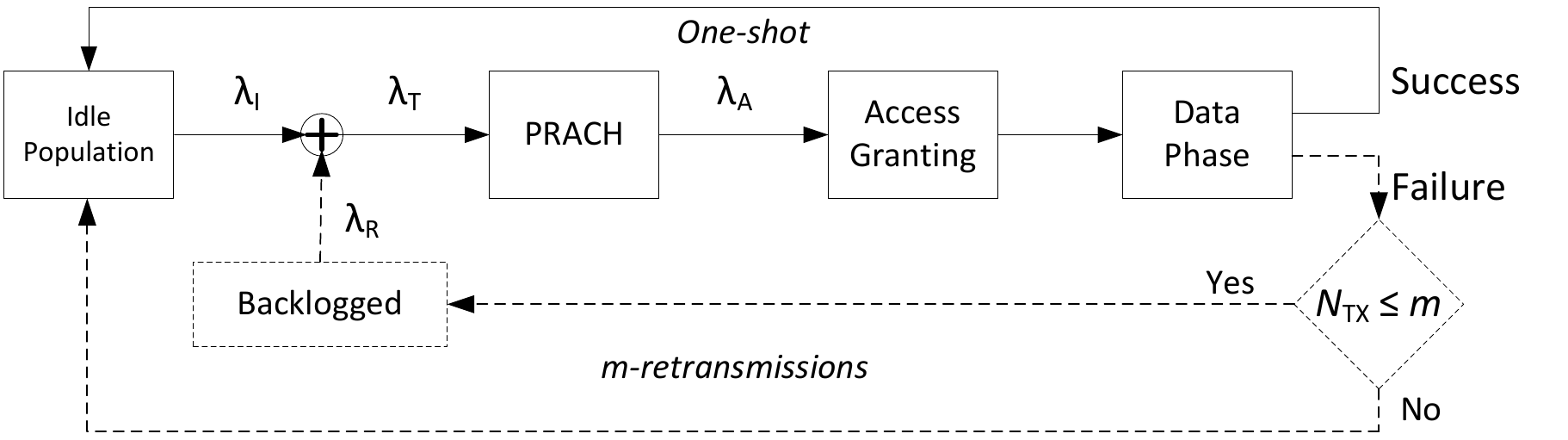}
    \protect\protect\protect\protect
    \caption{Flow diagram of LTE access reservation protocol: one-shot transmission model and full $m$-retransmissions model (dashed lines).}
    \label{fig:LTE_diagram}
\end{figure*}
In particular, $\lI=N \cdot \lambda_\text{app}$, where $\lambda_\text{app}$ is the transmission
generation rate at each UE. 
For each new data transmission,
up to $m$ retransmissions are allowed, resulting in a maximum of $m\!+\!1$ allowed
transmissions.
When transmissions fail and retransmission occurs, then an
additional load is put on the access reservation protocol, since the backlogged retransmissions $\lR$ add to the total rate $\lT$.
The total rate $\lambda_\text{T}$ corresponds to the traffic generated by the preamble activations by UEs in the PRACH channel.
After the PRACH stage, the traffic represented by $\lambda_\text{A}$ corresponds to the detected preambles, where $\lambda_\text{A} \leq \lambda_\text{T}$ since in case of a preamble collision only 1 preamble is activated.


As shown in Fig.~\ref{fig:LTE_diagram}, we split the access reservation model
into two parts: (i) the one-shot transmission part in
Fig.~\ref{fig:LTE_diagram}(a) (solid lines only) that models the bottlenecks at each stage of the access
reservation protocol; (ii) the $m$-retransmission part in
Fig.~\ref{fig:LTE_diagram}(b) (dashed lines), where finite number of retransmissions and backoffs are
modeled.
The modeling approach used for the two parts is an extension of our preliminary work \cite{nielsen2015tractable}, by taking into account the details of PDCCH, PDSCH and PUSCH channels, as presented in the following text.


\subsection{One-Shot Transmission Model} 
\label{sub:one_shot_transmission_model}

We are interested in characterizing how often a transmission from a UE fails. This happens when the transmission is not successful in the preamble contention or during the access granting phases. Conversely, for successful transmission, the request from the UE must not experience a preamble collision and there needs to be sufficient resources in the PDCCH, PDSCH, and PUSCH for the required messages. We model this as a sequence of two independent events:
\begin{equation}\label{eq:p_s_one-shot}
  p_\text{f}(\lT) = 1-\Big(1-p_\text{c}(\lT)\Big)\Big(1-p_\text{e}(\lT)\Big),
\end{equation}
where $p_\text{f}(\lT)$ is the probability of a failed UE transmission, 
$p_\text{c}(\lT)$ is the collision probability in the preamble
contention phase given a UE request rate $\lT$, and $p_\text{e}(\lT)$
is the probability of failure due to starvation of resources in the
PDCCH, PDSCH, or PUSCH.

\subsubsection{Preamble Contention Phase} 
\label{sub:preamble_contention_phase}
We start by computing $p_\text{c}(\lT)$. Let $d$ denote the number of
available preambles ($d=54$). Let the probability of not selecting the
same preamble as one other UE be $1-\frac{1}{d}$. Then the probability
of a UE selecting a preamble that has been selected by at least one
other UE given at $N_\text{T}$ contending UEs, is:
\begin{equation}
  \mathbb{P} \left(\mbox{Collision}|N_\text{T}\right) = 1 - \left( 1 - \frac{1}{d}\right)^{N_\text{T}-1}.
\end{equation}
Assuming Poisson arrivals with rate $\lT$, then:
\begin{align}\label{eq:p_c}
  p_\text{c}(\lambda_\text{T}) &=\!\sum_{i=1}^{+\infty}\!\left[ 1 - \left(\! 1 - \frac{1}{d}\right)^{i-1} \! \cdot \mathbb{P}(N_\text{T}\!=\!i , \lambda_\text{T} \cdot \delta_\text{RAO})\right] \\ \nonumber
          & \leq  1 - \left( 1 - \frac{1}{d}\right)^{\lambda_\text{T} \cdot \delta_\text{RAO}-1},
\end{align}
where $\mathbb{P}(N_\text{T} = i , \lambda_\text{T} \cdot \delta_\text{RAO})$ is the probability mass function of the Poisson distribution with arrival rate $\lambda_\text{T} \cdot \delta_\text{RAO}$. The inequality comes from applying Jensen's inequality~\cite{Gradshteyn2000} to the concave function $1 - \left( 1 - 1/d\right)^x$, where $\lambda_\text{T}$ is the total arrival rate (including retransmissions), and $\delta_\text{RAO}$ is the average number of subframes between RAOs.\footnote{E.g., $\delta_\text{RAO}\!=\!1$ if 10 RAOs per frame and $\delta_\text{RAO}\!=\!5$ if 2 RAOs per frame.}
The computed $p_\text{c}(\lambda_\text{T})$ is thus an upper bound on the collision probability.


\subsubsection{Access Granting Phase} 
\label{sub:access_grant_phase}

The mean number of activated preambles in the contention phase per RAO, is given by $\lA$.
As discussed in Section~\ref{sec:lte_overview}, we assume that the eNodeB is unable to discern between preambles that have been activated by a single user and multiple users, respectively.
This will lead to a higher $\lA$, than in the case where the eNodeB is able to detect the preamble collisions.
The main impact of this assumption is that there will be an increased rate of access granted requests, even though part of these correspond to collided preambles, which even if accepted will lead to retransmissions.
In addition to the rate of activated preambles $\lA$, we also need the rate of singletons, i.e., non-collided, successful preamble activations denoted by $\lS$.

The $\lA$ and $\lS$ can be well approximated, while assuming that the selection of each preamble by the contending users is independent, by,
\begin{align}\label{eq:lambda_t_nodetect}
    \lA =& \left[1 - \mathbb{P} \left(X=0 \right)\right] \cdot d/\delta_\text{RAO},\\
    \lS =& \mathbb{P} \left(X=1 \right) \cdot d/\delta_\text{RAO},
\end{align}
where $\mathbb{P}(X=k)$ is the probability of k successes, which can be well approximated with a Poisson distribution with arrival rate $\lambda_\text{pre}=\lT\delta_\text{RAO}/d$, i.e.:
\begin{equation}
    \mathbb{P} \left(X=k \right) \approx \frac{(\lambda_\text{pre})^k e^{-\lambda_\text{pre}}}{k!}.
\end{equation}

%

Since the limitations in the AG phase are primarily given from the
demanded resources from each of the channels PDCCH, PDSCH, and PUSCH
and the corresponding timing requirements, we assume that each of these
can be modeled as a separate queue with impatient costumers. That is,
we assume that the loss probability $p_\text{q}(\lA)$ can be seen as
the \emph{long-run fraction of costumers that are lost} in a queuing
system with impatient costumers \cite{de1985queueing}.

Based on the message exchange diagram in Section \ref{sub:lte_access_reservation_protocol}, we specify in the following text the arrival rate, service rate and the maximum latency for each of the channels PDCCH, PDSCH, and PUSCH.

In general, since LTE uses fixed size time slots, the most obvious
approach would be to use an M/D/1 model structure where service times are deterministic, as presented in
\cite{de1985queueing}. Unfortunately, the expression to compute the
fraction of lost customers $p_\text{q}(\lambda,\mu,T_\text{d})$ for the
M/D/1 queue does not have a closed form solution. However, the
equivalent expression for the M/M/1 queue, which assumes exponential duration service intervals, does have a closed form solution.
Through an extensive study, we have found that with the parameter
ranges that we use, there is no noticeable difference in the results. Furthermore, and most importantly, our results with this model fit well to simulation results, as shown in sec. \ref{sub:numerical_results}.
Thus, in the following we use the M/M/1 model to compute
$p_\text{q}(\lambda,\mu,T_\text{d})$ as:
\begin{align}
	p_\text{q}(\lambda,\mu,T_\text{d}) &= \frac{(1-\rho) \cdot \rho \cdot \Omega}{1-\rho^2 \cdot \Omega}, \text{ with } \Omega = e^{-\mu \cdot (1-\rho) \cdot \tau_\text{q}},
\end{align}
where $\rho=\frac{\lambda}{\mu}$ is the queue load, $\mu$ is the service rate, with $\tau_\text{q} = T_\text{d}-\frac{1}{\mu}$ and $T_\text{d}$ is the max waiting time.

Assuming we can use the M/M/1 model structure to obtain the failure
probabilities of the PDCCH, PDSCH, and PUSCH, we define
$p_\text{e}(\lT)$ from \eqref{eq:p_s_one-shot} as:
\begin{align}\label{eq:p_e}
  p_\text{e}(\lT) = 1-
  &\Big(1-p_\text{q}(\lambda_\text{PDCCH},\mu_\text{PDCCH},T_\text{d-PDCCH})\Big)\nonumber\\
  \cdot &\Big(1-p_\text{q}(\lambda_\text{PDSCH},\mu_\text{PDSCH},T_\text{d-PDCSH})\Big) \nonumber\\
  \cdot &\Big(1-p_\text{q}(\lambda_\text{PUSCH},\mu_\text{PUSCH},T_\text{d-PUSCH})\Big),
\end{align}
where the respective $\lambda$, $\mu$, and $T_\text{d}$ values are derived in the following. For the $\lambda$ values, we elaborate in Table \ref{tab:channel_lambdas} the amount of resources used in each of the channels for the relevant messages from Table \ref{tab:signalingMessages}.
For each, the resources are given in terms of PDCCH, PDSCH, or PUSCH elements per subframe. The model parameters are described in Table \ref{tab:queue_model_variables}.

The used M/M/1 model requires a single timeout value to specify the impatience threshold of the costumers. However, in the modeled LTE access procedure, there are several timers involved that cover different and sometimes overlapping parts of the message exchange. While this clearly cannot be modeled very accurately with the M/M/1 model used here, we will simply use a typical minimum timer value for each of the channels. Assuming that LTE has not been designed with timer values so low that the capacity is limited by timeouts and not by resource scarcity, this simplifying assumption should not have any significant impact on the results.


\setlength\tabcolsep{3 pt}
\begin{table*}[bt]
\centering\footnotesize
\caption{Amount of channel resources used for PDCCH, PDSCH and PUSCH channels. For short message format, only bold messages are used (RAR, RRC req., RRC comp., and data).}
\label{tab:channel_lambdas}
\begin{tabular}{cc|c|c|c|c|c|c|c|c|c}
\multicolumn{1}{l|}{{\it }}               				&\multicolumn{3}{c|}{{\it ARP}}   & \multicolumn{6}{c|}{{\it Additional signaling}} & \multicolumn{1}{l}{{\it }}                 \\
\multicolumn{1}{r|}{}                     				& \multirow{2}{*}{\bf RAR} 			& \multirow{2}{*}{\begin{minipage}[t]{1.1cm}{\bf RRC Request}\end{minipage}} & \multirow{2}{*}{\begin{minipage}[t]{1.1cm}{\it RRC Connect}\end{minipage}} & \multirow{2}{*}{\begin{minipage}[t]{1.35cm}{\bf RRC Complete}\end{minipage}} & \multirow{2}{*}{\begin{minipage}[t]{1.1cm}{\it Reconf. DL}\end{minipage}}  & \multirow{2}{*}{\begin{minipage}[t]{1.1cm}{\it Reconf. UL}\end{minipage}} & \multirow{2}{*}{\begin{minipage}[t]{1.1cm}{\it Security Cmd.}\end{minipage}} & \multirow{2}{*}{\begin{minipage}[t]{1.1cm}{\it Security Config.}\end{minipage}} & \multirow{2}{*}{\begin{minipage}[t]{1.25cm}{\it Security Complete}\end{minipage}} & \multirow{2}{*}{\begin{minipage}[t]{1.1cm}{\bf Data}\end{minipage}}            \\
\multicolumn{1}{r|}{{\it Channel}}                     &                                    & 		                                      &                                      &                                      &                                    &                                &                                          &         &                                     &            \\ \hline
\multicolumn{1}{r|}{{\it PDCCH}}                           & $\frac{1-e^{-\lT \delta_\text{RAO}}}{\delta_\text{RAO}}$                              & 0                                               & $\lS$                             & $\lS$                             & $\lS$                               & $\lS$                               & $\lS$                                & $\lS$ & 0                                                & \multicolumn{1}{c}{$\ceil{\frac{B_\text{data}}{N_\text{frag} B_\text{RB}}}$} \\
\multicolumn{1}{r|}{{\it PDSCH}}                           & $\ceil{\lA \frac{B_\text{RAR}}{B_\text{RB}}}$ & 0                                       & $\lS \ceil{\frac{B_\text{conn}}{B_\text{RB}}}$ & 0                                        & $\lS \ceil{\frac{B_\text{r-DL}}{B_\text{RB}}}$ & 0                                          & $\lS \ceil{\frac{B_\text{s-cmd}}{B_\text{RB}}}$ & 0            & 0                                                & \multicolumn{1}{c}{0}                     \\
\multicolumn{1}{r|}{{\it PUSCH}}                           & 0                                               & $\lA \ceil{\frac{B_\text{req}}{B_\text{RB}}}$ & 0                                        & $\lS \ceil{\frac{B_\text{comp}}{B_\text{RB}}}$ & 0                                          & $\lS \ceil{\frac{B_\text{r-UL}}{B_\text{RB}}}$ & 0                                           & 0                    & $\lS \ceil{\frac{B_\text{comp}}{B_\text{RB}}}$ & \multicolumn{1}{c}{$\lS \ceil{\frac{B_\text{data}}{B_\text{RB}}}$}              \\
\end{tabular}
\end{table*}
\setlength\tabcolsep{6 pt} 

\paragraph{PDCCH model}
The arrival rate for the PDCCH model $\lambda_\text{PDCCH}$, which describes the number of used PDCCH elements per subframe, is given as the sum of the PDCCH row in Table \ref{tab:channel_lambdas}. The service rate $\mu_\text{PDCCH}$ is the number of available PDCCH slots per subframe, i.e., $N_\text{PDCCH}$, and the timer value is the standard RAR timeout:
\begin{align*}
 	\lambda_\text{PDCCH} &= \frac{1-e^{-\lT \cdot \delta_\text{RAO}}}{\delta_\text{RAO}} \!+\!\lS (6 + \ceil[\bigg]{\frac{B_\text{data}}{N_\text{frag} B_\text{RB}}}) \\ \nonumber
 	\mu_\text{PDCCH} &= N_\text{PDCCH} \\ \nonumber
 	T_\text{d-PDCCH} &= 10,
\end{align*}
where $\ceil{x}$ is the smallest integer not less than $x$.

\paragraph{PDSCH model}
Similarly, the arrival rate for the PDSCH model is the sum of the corresponding row in Table \ref{tab:channel_lambdas}, the service rate is the number of available PDSCH elements per subframe, and the timer value is set to 40, which is a typical minimum value of the PDSCH related timers.

\begin{align*}
	\lambda_\text{PDSCH}\!&=\!\ceil[\bigg]{ \frac{\lA B_\text{RAR}}{B_\text{RB}} }\!+\!
    \lS\!\left(\!\ceil[\bigg]{\frac{B_\text{conn}}{B_\text{RB}}}\!+\!
 	\ceil[\bigg]{\frac{B_\text{r-DL}}{B_\text{RB}}}\!+\!
 	\ceil[\bigg]{\frac{B_\text{s-cmd}}{B_\text{RB}}}\!\right) \\ \nonumber
 	\mu_\text{PDSCH} &= N_\text{DLRB} \\ \nonumber
 	T_\text{d-PDSCH} &= 40.
\end{align*}

\paragraph{PUSCH model}
Finally, as above, the arrival rate for the PUSCH model is the sum of the corresponding row in Table \ref{tab:channel_lambdas}, the service rate is the number of available PUSCH elements per subframe subtracted the resources used for RAOs, and the timer value is set to 40, which is a typical minimum value of the PUSCH related timers.

\begin{align*}
	\lambda_\text{PUSCH} = &\lA\!\ceil[\bigg]{\frac{B_\text{req}}{B_\text{RB}}}+\lS\!\left( \ceil[\bigg]{\frac{B_\text{comp}}{B_\text{RB}}}\!+\!
 	\ceil[\bigg]{\frac{B_\text{r-UL}}{B_\text{RB}}}\!+\!
 	\ceil[\bigg]{\frac{B_\text{s-comp}}{B_\text{RB}}}\!+\!
 	\ceil[\bigg]{\frac{B_\text{data}}{B_\text{RB}}} \right) \\
 	\mu_\text{PUSCH} = &N_\text{ULRB}-6 \cdot \frac{10}{\delta_\text{RAO}}\\
 	T_\text{d-PUSCH} = &40.
\end{align*}

\begin{table}[tb]
	\centering\footnotesize
	\caption{Variable definitions}
	\label{tab:queue_model_variables}
	\begin{tabular}{|l|c|l|}
	\hline
	\textbf{Variable} & \textbf{Value} & \textbf{Description} \\ \hline
		$B_\text{RAR}$ & $ 8 $ & number of bytes used for the RAR message \\
		$B_\text{RB}$ & $ 36 $ & number of bytes per resource block\\
		$B_\text{req}$ & $ 7 $ & size of RRC request message in bytes \\
		$B_\text{conn}$ & $ 38 $ & size of RRC connect message in bytes \\
		$B_\text{comp}$ & $ 20 $ & size of RRC complete message in bytes \\
		$B_\text{r-DL}$ & $ 118 $ & size of RRC reconfigure DL message in bytes \\
		$B_\text{r-UL}$ & $ 10 $ & size of RRC reconfigure UL message in bytes \\
		$B_\text{s-cmd}$ & $ 11 $ & size of security command message in bytes \\
		$B_\text{s-comp}$ & $ 13 $ & size of security complete message in bytes \\
		$B_\text{data}$ & Variable & size of the data payload in bytes\\
		$N_\text{PDCCH}$ & Variable & number of PDCCH pointers per subframe\\
		$N_\text{DLRB}$ & Variable & number of resource blocks in PDSCH \\
		$N_\text{ULRB}$ & Variable & number of resource blocks in PUSCH \\
		$N_\text{frag}$ & 6 & fragmentation threshold in RBs\\
		\hline
	\end{tabular}
\end{table}

\subsection{m-Retransmissions Model} 
\label{sub:t_retransmissions_model}

During the ARP, UEs may experience failures of the transmitted packets (MSG1
and MSG3) and the received packets (MSG2 and MSG4). When a failure occurs with
probability $p_\text{f}(\lT)$, the total arrival rate $\lT$ changes to
represent also the additional arrivals of retransmissions. Further,
these additional arrivals affect the probability of failure again. To
model this behavior, we apply the two-dimensional Markov chain approach
first presented in \cite{bianchi2000performance}. The LTE adapted version
of this model have already been proposed in
\cite{nielsen2015tractable,yang2012performance} and in this work we
consider an extended version of the model in
\cite{nielsen2015tractable} to explicitly model the transmissions of
MSG1 and MSG3.

\begin{figure*}[t]
  \centering \includegraphics[width=13cm]{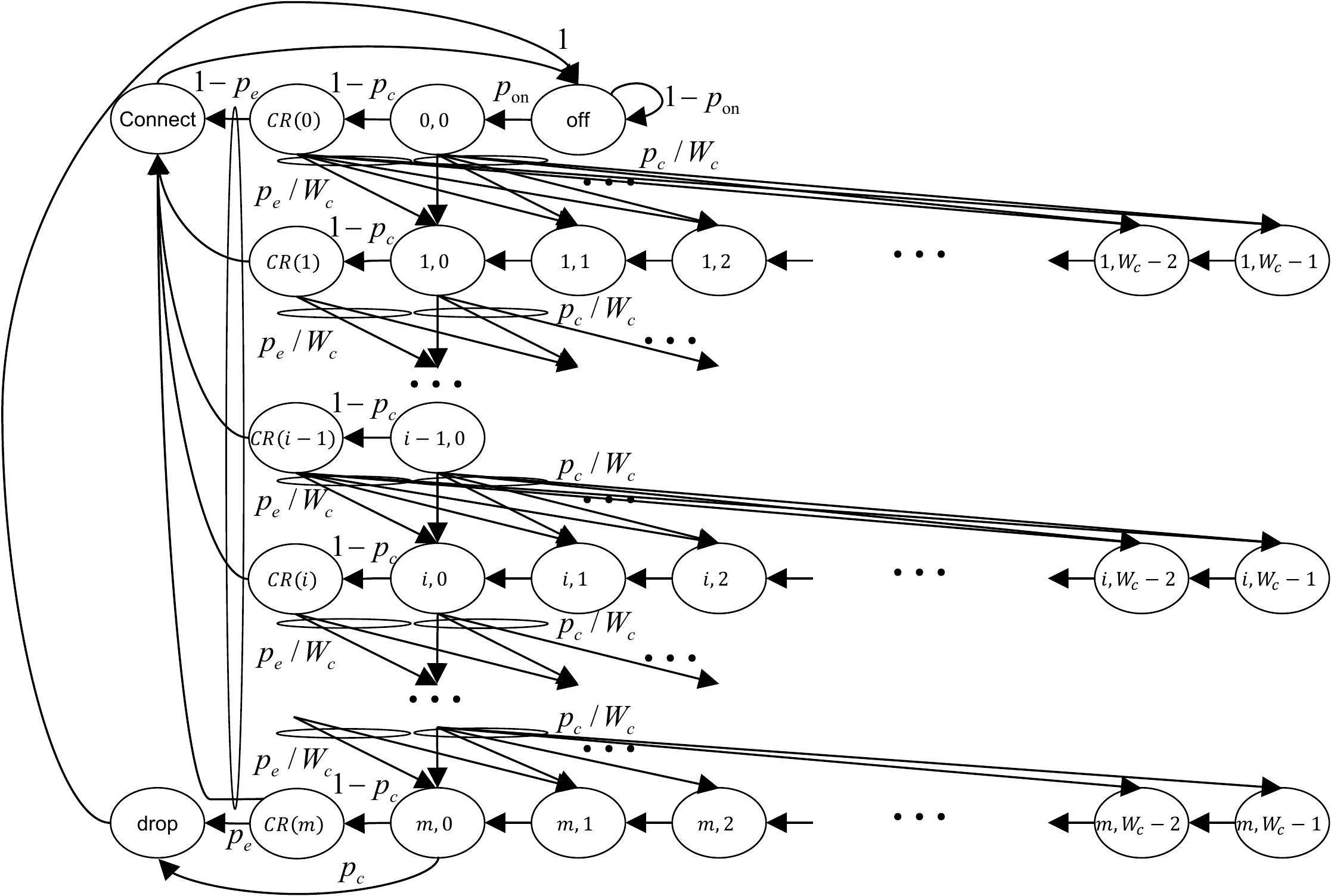}
  \caption{Markov chain model for $m$ retransmissions during the ARP.}
  \label{fig:mc_retx_model}
\end{figure*}

Fig.~\ref{fig:mc_retx_model} shows the structure of the Markov chain
model for $m$ retransmissions during the ARP. The uplink traffic at UE
is generated with probability $p_\text{on}$. The UE enters the
initial transmission state $\{0,0\}$ from the \emph{off} state:
\begin{equation*}
  \mathbb{P}\left( {\left. {0,0} \right|{\text{off}}} \right) = {p_\text{on}},
\end{equation*}
where $p_\text{on}$ is the traffic generation probability defined as
$p_\text{on} = 1-e^{-\lambda_{\text{I}}}$.

The state depicted $\{i, k\}$ represents the $i$th preamble
retransmission attempt and $k$th backoff counter. Retransmission
attempts are allowed up to $m$ times. The maximum backoff window size is
denoted by $W_c$. If a preamble transmission is not successful, the backoff
counter is increased and a random backoff state is entered with probability:
\begin{equation*}
  \mathbb{P}\left( {\left. {i,k} \right|i - 1,0} \right) = \frac{p_c}{W_c},\quad0 \leq k \leq W_c - 1, 1 \leq i \leq m,
\end{equation*}
where $p_c$ denotes the collision probability of the preamble
transmission.

The \emph{CR(i)} state represents the connect request attempt after the
success of the $i$th preamble transmission attempt. The transition probability is:
\begin{equation*}
  \mathbb{P}\left( {\left. {CR(i)} \right|i,0} \right) = {1 - p_c},\quad0 \leq i \leq m.
\end{equation*}

If the connect request attempt succeeds, the UE will be in the
\emph{connect} state. The transition probability is:
\begin{equation*}
  \mathbb{P}\left( {\left. {\text{connect}} \right|CR(i)} \right) = {1 - p_e},\quad0 \leq i \leq m.
\end{equation*}
where $p_e$ denotes the error probability of the connection request.

If the connection request is unsuccessful, the backoff counter is also
increased:
\begin{equation*}
  \mathbb{P}\left( {\left. i,k \right|CR(i-1)} \right) = \frac{p_e}{W_c},\quad1 \leq i \leq m.
\end{equation*}

The UE enters the \emph{drop} state if all attempts of preamble
transmissions and resource requests are failed:
\begin{align*}
  \mathbb{P}\left( {\left. {{\text{drop}}} \right|m,0} \right) = p_c(\lambda_\text{T}),\\
  \mathbb{P}\left( {\left. {{\text{drop}}} \right|CR(m)} \right) = p_e(\lambda_\text{T}).
\end{align*}

The UE will always return to the \emph{off} state after the \emph{connect} or
the \emph{drop} states, i.e., $\mathbb{P}\left( {\left.
{{\text{off}}} \right|{\text{drop}}} \right) = \mathbb{P}\left( {\left.
{{\text{off}}} \right|{\text{connect}}} \right) = 1$.

Let $b_{CR(i)}$, $b_{i,k}$, $b_{\text{connect}}$, $b_{\text{drop}}$,
and $b_{\text{off}}$ be the steady state probability that a UE is at
states \emph{CR(i)}, $\{i,k\}$, \emph{connect}, \emph{drop}, and
\emph{off}, respectively. Then,
\begin{align*}
b_\text{off} = p_\text{on}b_\text{off} + b_\text{connect} + b_\text{drop}.
\end{align*}

The steady state probability $b_{i,0}$ is expressed as:
\begin{align}\label{eq:bi0}
  {b_{i,0}} &= {p_e}{b_{CR\left( {i - 1} \right)}} + {p_c}{b_{i - 1,0}} \nonumber\\
   &= {p_e}\left( {1 - {p_c}} \right){b_{i - 1,0}} + {p_c}{b_{i - 1,0}} \nonumber\\
   &= \left( {{p_e}\left( {1 - {p_c}} \right) + {p_c}} \right){b_{i - 1,0}} \nonumber\\
   &= {\left( {{p_e}\left( {1 - {p_c}} \right) + {p_c}} \right)^i}{b_{0,0}}.
\end{align}

Using \eqref{eq:bi0}, the steady state probability $b_{CR(i)}$ is
derived as:
\begin{align}\label{eq:bcr}
  {b_{CR\left( i \right)}} &= \left( {1 - {p_c}} \right){b_{i,0}} \nonumber\\
   &= \left( {1 - {p_c}} \right){\left( {{p_e}\left( {1 - {p_c}} \right) + {p_c}} \right)^i}{b_{0,0}} \nonumber\\
   &= \left( {1 - {p_c}} \right){\left( {{p_e}\left( {1 - {p_c}} \right) + {p_c}} \right)^i}{p_{{\text{on}}}}{b_{{\text{off}}}}.
\end{align}

Using \eqref{eq:bcr}, $b_{i,k}$ is derived as:
\begin{align}\label{eq:bik}
  {b_{i,k}} &= \frac{{{W_c} - k}}{{{W_c}}}\left( {{p_c}{b_{i - 1,0}} + {p_e}{b_{CR\left( {i - 1} \right)}}} \right) \nonumber\\
   &= \frac{{{W_c} - k}}{{{W_c}}}\left( {p_c}{{\left( {{p_e}\left( {1 - {p_c}} \right) + {p_c}} \right)}^{i - 1}}{b_{0,0}} +\right.\nonumber\\
   &\qquad\qquad\left.{p_e}\left( {1 - {p_c}} \right){{\left( {{p_e}\left( {1 - {p_c}} \right) + {p_c}} \right)}^{i - 1}}{b_{0,0}} \right) \nonumber\\
   &= \frac{{{W_c} - k}}{{{W_c}}}{\left( {{p_e}\left( {1 - {p_c}} \right) + {p_c}} \right)^i}{p_{{\text{on}}}}{b_{{\text{off}}}},
\end{align}
for $1 \leq i \leq m$ and $0 \leq k \leq W_\text{c}-1$.

Using \eqref{eq:bcr}, $b_\text{connect}$ and $b_\text{drop}$ are
derived as:
\begin{align}\label{eq:bconn}
  {b_{{\text{connect}}}} &= \sum\limits_{i = 0}^m {\left( {1 - {p_e}} \right){b_{CR\left( i \right)}}}  \nonumber\\
   &= \sum\limits_{i = 0}^m {\left( {1 - {p_e}} \right)\left( {1 - {p_c}} \right){{\left( {{p_e}\left( {1 - {p_c}} \right) + {p_c}} \right)}^i}{p_{{\text{on}}}}{b_{{\text{off}}}}}  \nonumber\\
   &= \left( {1 - {p_e}} \right)\left( {1 - {p_c}} \right){p_{{\text{on}}}}{b_{{\text{off}}}}\frac{{1 - {{\left( {{p_e}\left( {1 - {p_c}} \right) + {p_c}} \right)}^{m + 1}}}}{{1 - \left( {{p_e}\left( {1 - {p_c}} \right) + {p_c}} \right)}} \nonumber\\
   &= \left( {1 - {{\left( {{p_e}\left( {1 - {p_c}} \right) + {p_c}} \right)}^{m + 1}}} \right){p_{{\text{on}}}}{b_{{\text{off}}}},
\end{align}
\begin{align}\label{eq:bdrop}
  {b_{{\text{drop}}}} &= {p_e}{b_{CR\left( m \right)}} + {p_c}{b_{m,0}} \nonumber\\
   &= {p_e}\left( {1 - {p_c}} \right){\left( {{p_e}\left( {1 - {p_c}} \right) + {p_c}} \right)^m}{p_{{\text{on}}}}{b_{{\text{off}}}} +\nonumber\\
   &\qquad\qquad {p_c}{\left( {{p_e}\left( {1 - {p_c}} \right) + {p_c}} \right)^m}{p_{{\text{on}}}}{b_{{\text{off}}}} \nonumber\\
   &= {\left( {{p_e}\left( {1 - {p_c}} \right) + {p_c}} \right)^{m + 1}}{p_{{\text{on}}}}{b_{{\text{off}}}}.
\end{align}

By imposing the probability normalization condition
\begin{align*}
1 = {b_{{\text{off}}}} + {b_{{\text{connect}}}} + {b_{{\text{drop}}}} + {b_{0,0}} + \sum\limits_{i = 1}^m {\sum\limits_{k = 0}^{{W_c} - 1} {{b_{i,k}}} }  + \sum\limits_{i = 0}^m {{b_{CR\left( i \right)}}},
\end{align*}
we find ${b_{{\text{off}}}}$ as:

\ifCLASSOPTIONtwocolumn
  \begin{strip}
    \begin{equation}
      {b_{{\text{off}}}} \!=\! \frac{{2( {1 - {p_e}} )\!( {1 - {p_c}} )}}{{2( {1 \!+\! 2{p_{{\text{on}}}}} )\!( {1 \!-\! {p_e}} )\!( {1 \!-\! {p_c}} ) \!+\! ( {{W_c} \!+\! 1} ){p_{{\text{on}}}}( {{p_e}( {1 \!-\! {p_c}} ) \!+\! {p_c}} )\!( {1 \!-\! {{( {{p_e}( {1 \!-\! {p_c}} ) \!+\! {p_c}} )}^m}} ) \!+\! 2( {1 \!-\! {p_c}} ){p_{{\text{on}}}}( {1 \!-\! {{( {{p_e}( {1 \!-\! {p_c}} ) \!+\! {p_c}} )}^{m \!+\! 1}}} )}}
    \end{equation}
  \end{strip}
\else
  \begin{equation}
    \resizebox{\textwidth}{!}{$ {b_{{\text{off}}}} \!=\! \frac{{2( {1 - {p_e}} )\!( {1 - {p_c}} )}}{{2( {1 \!+\! 2{p_{{\text{on}}}}} )\!( {1 \!-\! {p_e}} )\!( {1 \!-\! {p_c}} ) \!+\! ( {{W_c} \!+\! 1} ){p_{{\text{on}}}}( {{p_e}( {1 \!-\! {p_c}} ) \!+\! {p_c}} )\!( {1 \!-\! {{( {{p_e}( {1 \!-\! {p_c}} ) \!+\! {p_c}} )}^m}} ) \!+\! 2( {1 \!-\! {p_c}} ){p_{{\text{on}}}}( {1 \!-\! {{( {{p_e}( {1 \!-\! {p_c}} ) \!+\! {p_c}} )}^{m \!+\! 1}}} )}} $}.
  \end{equation}
\fi

Using $b_\text{connect}$ and $b_\text{drop}$, the outage probability
can be computed as:
\begin{align}
  P_\text{outage} = \frac{b_\text{drop}}{b_\text{drop}+b_\text{connect}} = (p_\text{e}(1-p_\text{c})+p_\text{c})^{m + 1}. \label{eq:p_outage_mc}
\end{align}

Rearranging the base of the exponentiation in \eqref{eq:p_outage_mc} gives:
\begin{align}
p_\text{e}(1-p_\text{c})+p_\text{c} = 1-(1-p_\text{e})(1-p_\text{c})=p_\text{f},
\end{align}
and thus we have that
\begin{align}
  P_\text{outage} = p_\text{f}^{m + 1}.
\end{align}
\textcolor{blue}{Hereby, the derivation of $P_\text{outage}$ can model the failure of a connection request message; note that the derivation of $P_\text{outage}$ in \cite{nielsen2015tractable} assumed that the connection request message is always delivered successfully.} Additionally, this means that the connect request failures can justifiably be assumed to be independent from the preamble collisions, as assumed for the one-shot transmission model in \eqref{eq:p_s_one-shot}, and as we assumed in \cite{nielsen2015tractable} where eq. \eqref{eq:p_s_one-shot} was used for the m-retransmissions model as well.

Further, as shown in \cite{nielsen2015tractable}, the number of required
transmissions can be approximated from the number of failures:
\begin{align}\label{eq:entx}
  N_\text{TX}(\lambda_\text{T}) &= \sum\limits_{i = 0}^{m} {p_\text{f}^i} = \frac{{1 - p_\text{f}^{m + 1}}}{{1 - {p_\text{f}}}}.
\end{align}

Using \eqref{eq:entx}, the value of $\lambda_\text{T}$ can be obtained
by solving the following iterative equation:
\begin{align}
\lambda_\text{T} = N_\text{TX}(\lambda_\text{T}) \cdot \lambda_\text{I}
= \lambda_\text{I} \frac{{1 - {{\left( {{p_e}\left( {1 - {p_c}} \right) + {p_c}} \right)}^{m + 1}}}}{{1 - \left( {{p_e}\left( {1 - {p_c}} \right) + {p_c}} \right)}},
\end{align}
where $\lI$ is constant but $p_\text{c}$ and $p_\text{e}$ are both functions of $\lT$ as defined in \eqref{eq:p_c} and \eqref{eq:p_e}.


\section{System Performance Evaluation} 
\label{sec:results}

In this section we first describe the traffic models used here. Thereafter we present and discuss numerical results, where we compare results from our analytical model to the simulation results.

\subsection{Model of the Smart Grid Traffic} 
\label{sec:smart_grid_traffic}

At the time of writing, there is no standardized traffic model that could be used to describe reporting activities of the eSMs.
In the following, we develop a model by considering the typical smart metering traffic models and enhancing them in order to achieve PMU-like functionalities that eSMs are expected to have.

In the literature there are different examples of traffic models for smart meters, such as \cite{hossain2012smart,deshpande2011differentiated,khan2013comprehensive,6629817}.
Of these, the OpenSG \emph{Smart Grid Networks System Requirements Specification} (described in \cite{hossain2012smart}) from the Utilities Communications Architecture (UCA) user group is the most coherent and detailed network requirement specification.
This specification describes the typical configuration where billing reports are collected as often as every 1 hour for industrial smart meters and every 4 hours for residential smart meters. While this is sufficient for billing purposes, such low reporting frequency does not allow real-time monitoring and control. A way to enable this, as proposed and analysed in our work in \cite{nielsen2015magazine}, would be to drastically increase the reporting frequency of all smart meters so that reports are collected, e.g., every 10 seconds. While such a configuration is not described in OpenSG \cite{hossain2012smart}, it is mentioned that on-demand meter read response messages are 100 bytes, wherefore we will use this value in the following evaluation.

Besides the basic measurements of consumption and production, the distribution system operators need to collect more detailed information of the distribution grid behavior in the form of power phasors from certain, strategically chosen measurement points. As an example in the following numerical results, we assume that every 10 seconds an eSM sends a measurement report that consist of concatenated PMU measurements (1~Hz sample rate) from the preceding 10 second measurement interval. The samples are, as specified in PMU standards IEEE 1588 \cite{lee2005ieee} and C37.118 \cite{martin2008C37118}, timestamped using GPS time precision.
Assuming that the floating point PMU frame format from IEEE 1588 is used and that each sample covers 6 phasors, 1 analog value and 1 digital value, each PMU sample accounts to 76 bytes. Adding UDP header (8 bytes) and IPv6 header (40 bytes) to each report of 50 PMU samples, an eSM packet is 808 bytes. Assuming that additional headers, e.g., for security purposes are needed, we round this up to an assumed eSM packet size of 1000 bytes.


\subsection{Numerical Results} 
\label{sub:numerical_results}

	
	In order to evaluate the performance of the LTE system for smart metering and validate the proposed model, we have developed an event-driven simulator in MATLAB.
	This simulator models the main downlink and uplink channels.
	More specifically, we model the downlink control and data channels (PDCCH and PDSCH respectively); and the uplink data and random access channels (PUSCH and PRACH).
	The uplink control channel (PUCCH) can be shared among multiple users and its impact on the performance for typical configurations can be neglected \cite{signaling3GPP}.
	We consider a typical 5~MHz (25~RBs) cell configured with one RAO every 5~ms ($\delta_\text{RAO} = 5$), 54 available preambles ($d$) for contention and a backoff value of 20~ms \cite{typicalValues}.
	In addition, we also investigate the performance of the smallest bandwidth cell in LTE, which corresponds to a 1.4~MHz (6~RBs), where $\delta_\text{RAO} = 20$.
	Link adaptation is out of the scope of this paper and therefore we focus in the lowest modulation in LTE (QPSK).
	The packet fragmentation threshold $N_\text{frag}$ is set to 6~RBs, which corresponds to the maximum uplink bandwidth transmission foreseen for LTE-M (low cost LTE for M2M)~\cite{typicalValues,3GPPR13}.
	The maximum number of PRACH retransmissions for a given data packet is set to a typical value ($m=9$) \cite{typicalValues}.
	Further we consider SMs and eSMs reporting every 10~s, which allows for a more frequent monitoring of the grid \cite{nielsen2015magazine}.
	The report size is set to $RS = \{100, 1000\}$~bytes, which illustrates small and large payloads described in the previous section (one order of magnitude of difference) impact on the system performance.
	However, we note that the proposed model can be also used for different payloads sizes and reporting intervals.
	The rest of parameters of interest are listed in Table~\ref{tab:lte_system_parameters}.
	\begin{table}[t]
		\centering\footnotesize
		\caption{LTE simulation and model parameters}
		\begin{tabular}{l|c}
			\textbf{Parameter} & \textbf{Value} \\ \hline
				Preambles per RAO (d) & 54 \\
				Subframes between RAOs ($\delta_\text{RAO}$) & 20 or 5 \\
				Max number of retransmissions ($m$) & 0 or 9 \\
				CFI Value & 3 \cite{3GPPTS36.508} \\
				Number of CCEs ($\mu$) & 6 or 21 \\
				System bandwidth & 1.4~MHz or 5~MHz \\
				eNodeB processing time & 3~ms \\
				UE processing time & 3~ms \\
				MSG~2 window ($t_\text{RAR}$) & 10~ms \\
				Contention time-out ($t_{\mathrm{CRT}}$) & 40~ms \\
				Backoff limit ($W_\text{c}$) & 20~ms  \\
				Rest of Messages window & 40~ms
		\end{tabular}
		\label{tab:lte_system_parameters}
	\end{table}
	\begin{figure}[t]
    	\centering
    	\includegraphics[width=\figwidth]{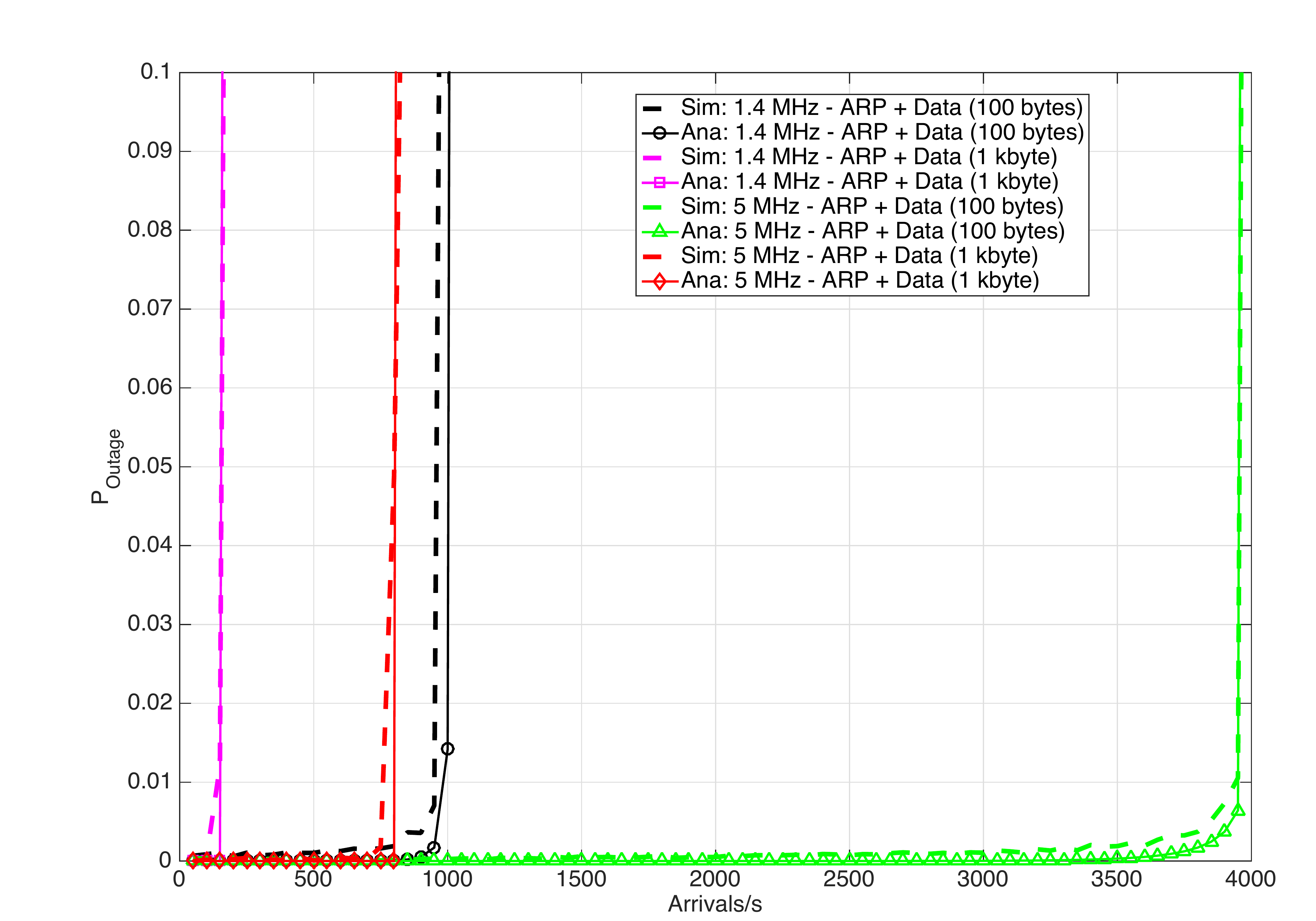}
    	\caption{Probability of outage in LTE with respect the number of M2M arrivals per second in a 1.4~MHz and 5~MHz system for different models, payloads and number of RAOs.}
    	\label{fig:compareSmallLarge}
	\end{figure}
	The evaluation is performed in terms of outage and number of supported users.
	\textcolor{blue}{The outage probability is defined as the probability of a device not being served before reaching the maximum number of PRACH transmissions and its corresponding analytical expression is given in \eqref{eq:p_outage_mc}.}
	
	First we consider the case where immediately after the ARP (i.e., after MSG~4), the data transmission starts. That is, we have only the messages shown in bold text in Table \ref{tab:channel_lambdas}\footnote{The case where the data transmission occurs immediately after the ARP, without the additional signaling denoted in Table~\ref{tab:signalingMessages}, is denoted as lightweight-signaling access and corresponds to an extreme case of signaling overhead reduction, beyond what has been proposed in 3GPP~\cite{3gpp2011tr37868,3gpp2013tr37869}.}.
	Fig.~\ref{fig:compareSmallLarge} shows the outage probability $P_{\text{outage}}$ for 1.4~MHz and 5~MHz systems, both for SM and eSM traffic models.
	It can be seen that the analytical model is very capable of capturing the outage point, where the system gets destabilized and the outage events become overwhelming. Since the intention is to characterize when the system is reliable, we focus on the region where the service outage is below 10\%.
	The impact of the payload (MAC layer limitations) becomes clear in Fig.~\ref{fig:compareSmallLarge}.
	A 1.4~MHz system can support a few hundreds (100 arrivals/s) for large eSM payloads (1000~bytes) and up to 1000 arrivals/s for small SM payloads (100~bytes).
	As expected, increasing the bandwidth does help to increase the capacity of the system, raising the number of supported arrivals to 700 arrivals/s and 4000 arrivals/s respectively.
	It should be noted that if the ARP is neglected and the focus is solely on the data capacity as in \cite{hagerling2014coverage,nist2011pap2}, up to 9000 arrivals/s can be supported.\textcolor{blue}{When compared to our results where the different ARP limitations are taken into account, it is clear that for M2M scenarios, data capacity based analyses are too simplistic and give overly optimistic results \cite{hagerling2014coverage,nist2011pap2}, which was also pointed out in \cite{Madueno2014}.}

	\begin{figure}[t]
    	\centering
    	\includegraphics[width=\figwidth]{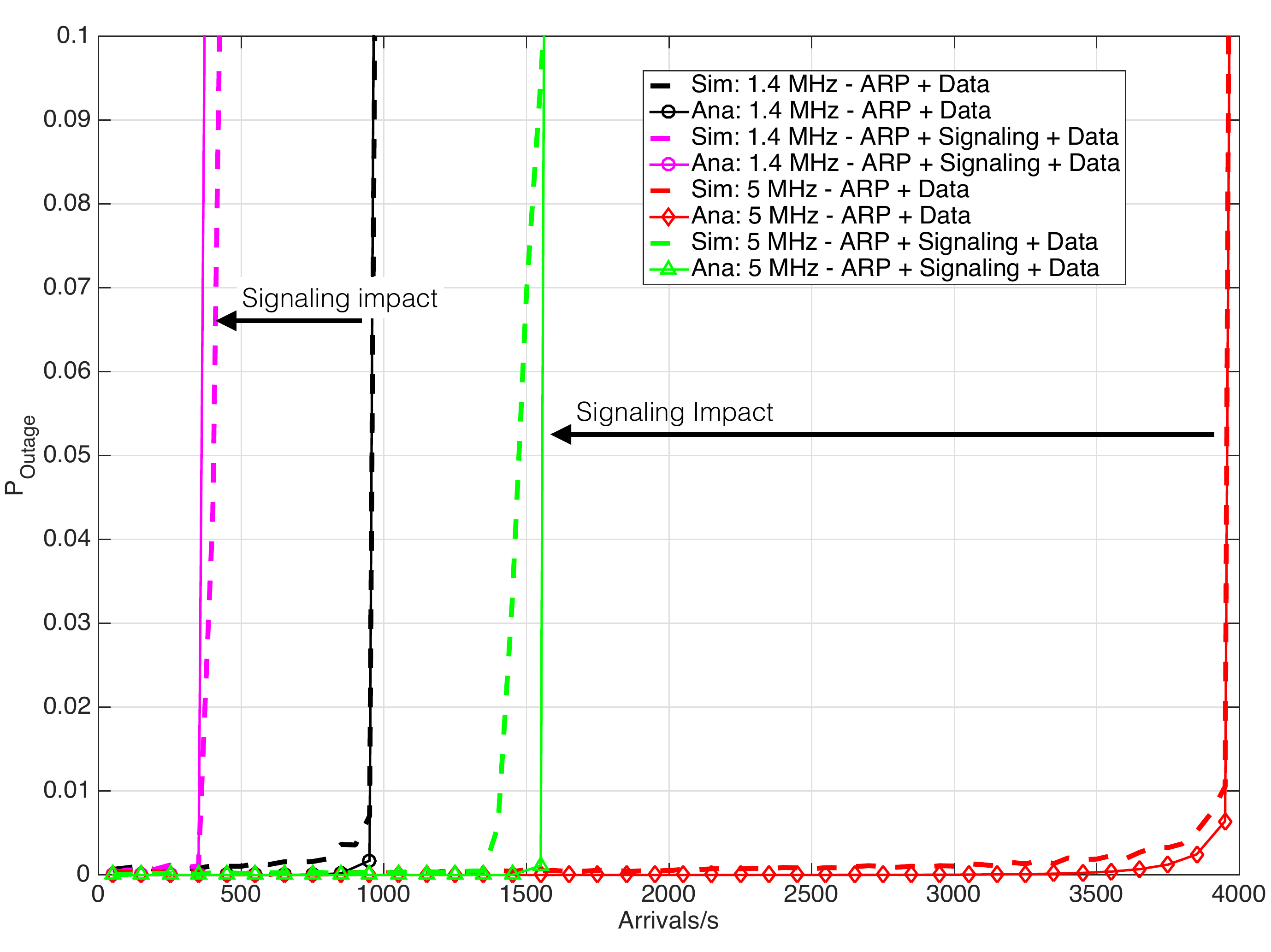}
    	\caption{Outage comparison for only ARP and data transmission (ARP + Data) and full message exchange (ARP + Signaling + Data).}
    	\label{fig:compareShortFull}
	\end{figure}

	In Fig.~\ref{fig:compareShortFull} we investigate the impact of the additional signaling messages that follows the ARP, as described in Section~\ref{sub:lte_access_reservation_protocol}.
	The striking conclusion is that, for both the 1.4~MHz and 5~MHz cases the number of supported arrivals is decreased by almost a factor of 3, decreasing from 1000 to 400 arrivals/s and from 4000 to 1500 arrivals/s respectively.
	Obviously, \emph{the additional signaling must be accounted for as it has a large impact on the system performance}.
	\begin{figure}[t]
    	\centering
    	\includegraphics[width=\figwidth]{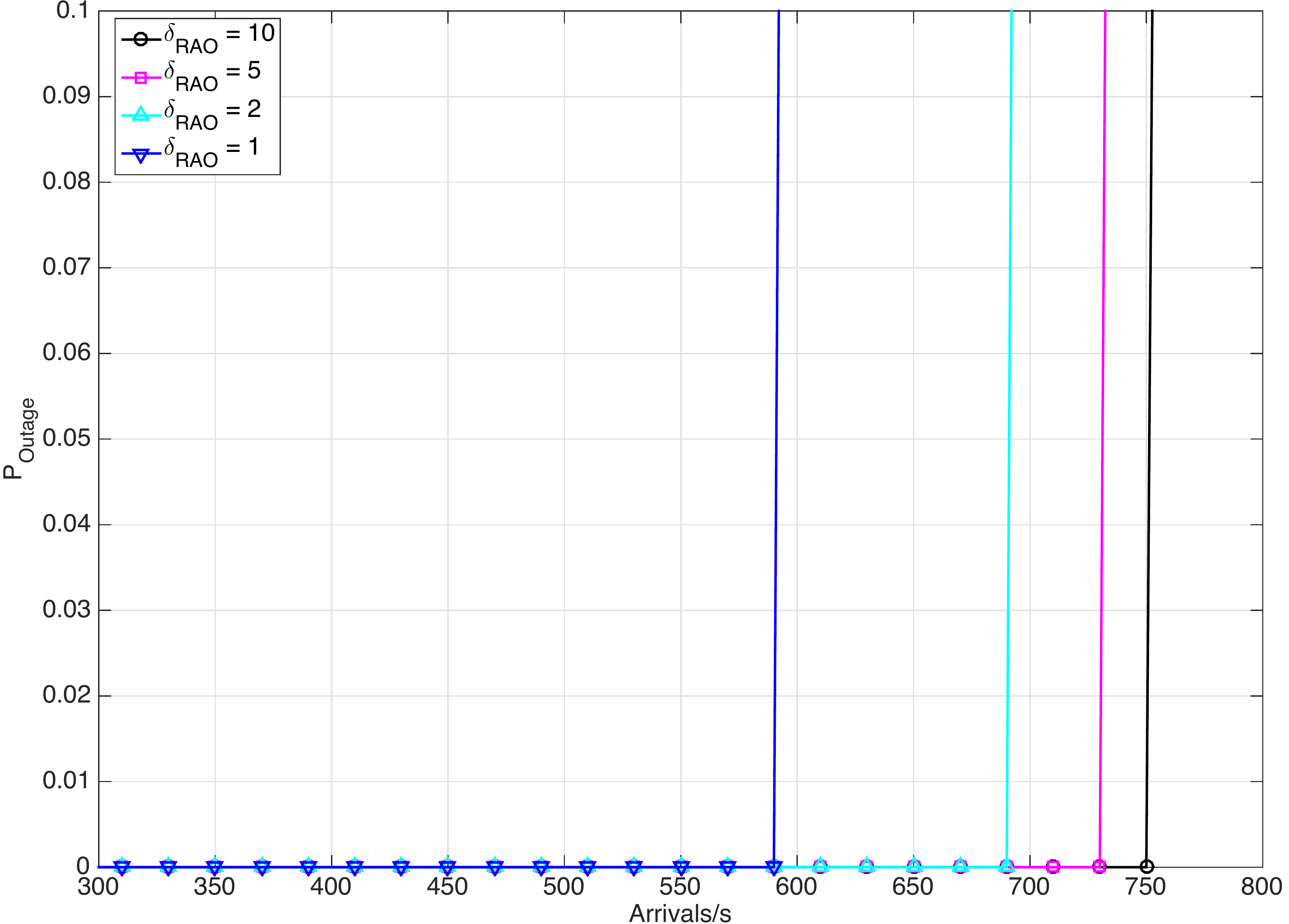}
    	\caption{Outage comparison for different number of RAOs per frame in a 5~MHz system with a payload of 1~kbyte (ARP + Signaling + Data).}
    	\label{fig:numberRAOs}
	\end{figure}

	Further, in Fig.~\ref{fig:numberRAOs} we illustrate the outage performance as the number of RAOs per frame is increased, i.e., when the distance between RAOs is decreased as $\delta_{\text{RAO}}=\{10, 5, 2, 1\}$ \textcolor{blue}{subframes} for the 5~MHz system with large payload and the entire sequence of messages considered.
	Although increasing the number of RAOs per frame is seen as the optimal solution for massive M2M \cite{R2104662}, it does not help when the rest of the limitations of the system is considered.
	It can be clearly seen that the best performance (supporting up to 750~arrivals/s) is achieved with a single RAO per frame ($\delta_{\text{RAO}}=10$), while the worst performance is present when the maximum number of RAOs per frame is selected ($\delta_{\text{RAO}}=1$).
	Similar behavior can be observed for other cases.
	\begin{figure}[t]
    	\centering
    	\includegraphics[width=\figwidth]{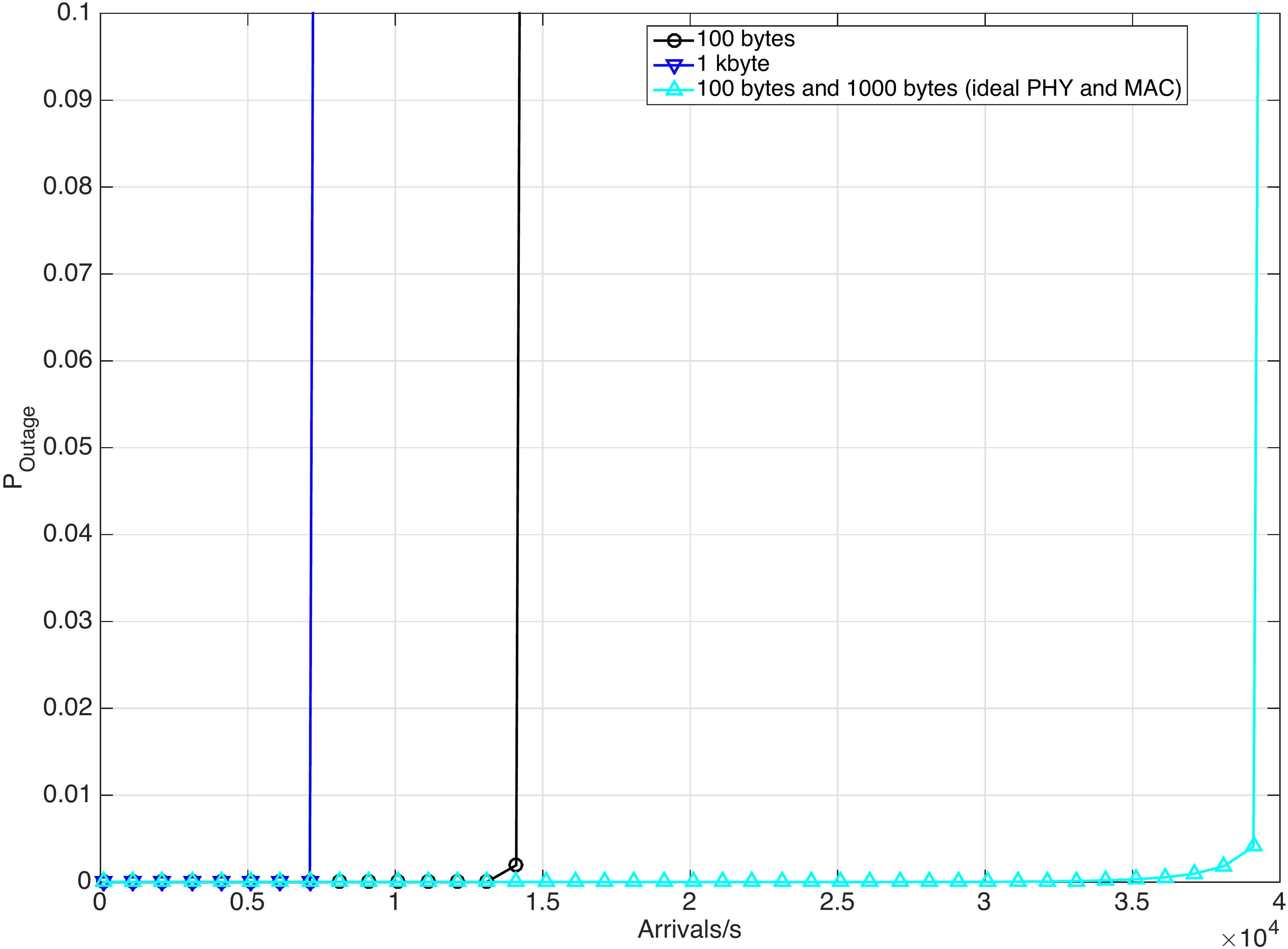}
    	\caption{Outage comparison for different number of RAOs per frame in a 5~MHz system with a payload of 1~kbyte (ARP + Signaling + Data).}
    	\label{fig:limitations}
	\end{figure}

	We conclude by illustrating the importance of considering not only the ARP limitations but also the PHY and MAC layer limitations in Fig.~\ref{fig:limitations}.
	The scenario considered is a 5~MHz system with 2~RAOs per frame ($\delta_{\text{RAO}}=5$) with 100~bytes and 1~kbyte.
	The 100~bytes case is limited by the number of PDCCH messages required, and therefore we see the outage peaks in approximately $1.5 \cdot 10^4$~arrivals/s.
	In the 1~kbyte case, the major limitation is the MAC layer, or more specifically the PUSCH, which limits the number of supported arrivals to $7000$~arrivals/s.
	It should be noted that the supported number of arrivals per second has been halved if the PUSCH limitation is considered.
	On the other hand, if we only consider the collisions in the PRACH we can support up to $3.9 \cdot 10^4$~arrivals/s, which represents an astonishing difference with respect to the actual performance of the system.
	
\section{Conclusion} 
\label{sec:conclusions}

One of the main messages brought by this paper is that the study of the performance of the LTE access in case of massive M2M traffic requires a fundamentally different approach compared to the study of human-type traffic. Specifically, in M2M, it is necessary to take into account the features of the actual channels used to exchange signaling information, such as PRACH, PDCCH and  PUSCH.
In case of small payloads, the main limitations are posed by PDCCH or PRACH if the system bandwidth is very large.
On the other hand, in case of larger payloads (1000 bytes), the limitations are posed by PUSCH.
Also, it was shown that, surprisingly, increasing the number of RAOs does not always help, as in most cases provision of RAOs per frame above a certain limit will negatively impact the performance.

While it is possible to obtain these results for any given scenario using tedious simulations, e.g., for different payload sizes or RAO configurations, we have shown that the analytical model developed in the paper, which can be rapidly implemented and evaluated, allows to obtain the service outage breaking point accurately.

The proposed modeling and evaluation of LTE access can be easily extended to include more limitations such as the PDSCH if the M2M service is also intensive in downlink messages.
However, judging from \cite{hossain2012smart} the downlink is barely used in smart grid monitoring applications, except for occasional software and firmware updates, and it is natural to assume that its impact can be neglected in such cases.

Another major insight is that the additional signaling that follows the ARP has very large impact on the capacity in terms of the number of supported devices; in the assessed setup we observed a reduction in the capacity by almost a factor of 3.
This calls for the consideration of a more efficient procedure in case of M2M connection establishment in future LTE standardization, e.g., a lightweight procedure in which the data report is sent immediately after the ARP.

We conclude by noting that, to the best of our knowledge, this the first study that accurately models and shows the full impact of the connection establishment on the support of massive M2M reporting in LTE, and, as such, may provide basis for the future standardization work.

\

\section*{Acknowledgment}
The research presented in this paper was partly funded by the EU
project SUNSEED, grant no. 619437, partly by the Danish Council for
Independent Research grant no. DFF-4005-00281 ``Evolving wireless
cellular systems for smart grid communications'', and partly supported
by the Danish High Technology Foundation via the Virtuoso project.

\bibliographystyle{IEEEtran}

\end{document}